\PassOptionsToPackage{final}{graphicx}
\documentclass[final, a4paper, UKenglish, cleveref, numberwithinsect, autoref]{lipics-v2021}
\pdfoutput=1 %
\nolinenumbers %
\hideLIPIcs %

\usepackage[utf8]{inputenc}
\usepackage[T1]{fontenc}
\usepackage{hyphenat}

\usepackage{ifdraft}
\usepackage{ifthen}
\usepackage{microtype}
\usepackage{amsthm}
\usepackage{thmtools}

\newboolean{proofsinappendix}
\setboolean{proofsinappendix}{true}
\ifdraft{}{%
  \setboolean{proofsinappendix}{true}
}

\newboolean{linktoappendix}
\setboolean{linktoappendix}{true}

\ifdraft{}{%
  \setboolean{linktoappendix}{false}
}

\usepackage{todonotes}
\usepackage{xspace}
\usepackage{mathtools}
\usepackage{stmaryrd}
\usepackage{etex}
\usepackage{environ}
\usepackage{etoolbox}
\usepackage{newfile}
\usepackage{enumitem}
\usepackage{hyperref}

\AtBeginDocument{
    \hypersetup{
        final,
        hidelinks,
        pdfsubject={},
    }
}

\usepackage[marginclue,footnote]{fixme}
\FXRegisterAuthor{tw}{atw}{TW}%

\presetkeys{todonotes}{
  color=yellow!20,
  bordercolor=yellow!80!black,
  linecolor=yellow,
  size=\scriptsize,
}{}

\newcommand{\descto}[3][]{\arrow[phantom]{#2}[#1]{\text{\footnotesize{}\begin{tabular}{c}#3\end{tabular}}}}
\tikzset{shiftarr/.style={
    rounded corners,%
    to path={--([#1]\tikztostart.center)
      -- ([#1]\tikztotarget.center) \tikztonodes
      -- (\tikztotarget)},
  }}

\tikzset{shiftarr/.style={
    rounded corners,%
    to path={--([#1]\tikztostart.center)
      -- ([#1]\tikztotarget.center) \tikztonodes
      -- (\tikztotarget)},
  }}

\usepackage{twshowkeys}

\newcommand{\resetCurThmBraces}{%
  \gdef\curThmBraceOpen{(}%
  \gdef\curThmBraceClose{)}}
\resetCurThmBraces
\newcommand{\removeThmBraces}{%
  \gdef\curThmBraceOpen{}%
  \gdef\curThmBraceClose{}}
\resetCurThmBraces

\newenvironment{notheorembrackets}{\removeThmBraces}{\resetCurThmBraces}

\usepackage{etoolbox}
\patchcmd{\thmhead}{(#3)}{\curThmBraceOpen #3\curThmBraceClose }{}{}

\newtheorem{notation}[theorem]{Notation}
\newtheorem{assumption}[theorem]{Assumption}
\newtheorem{construction}[theorem]{Construction}

\newcommand{\proofstep}[1]{\smallskip\noindent\textsf{\bfseries\color{lipicsGray}#1.}~}

\let\C\relax %
\let\G\relax %

\newcommand{\ToSet}{V}
\newcommand{\A}{\ensuremath{{\mathbb{A}}}\xspace}
\newcommand{\Q}{\ensuremath{{\mathbb{Q}}}\xspace}
\newcommand{\G}{\ensuremath{{\mathcal{G}}}\xspace}

\newcommand{\Perm}{\ensuremath{\mathfrak{S}_\mathsf{f}}}

\newcommand{\Aut}{\ensuremath{\mathsf{Aut}}}
\newcommand{\Fin}{\ensuremath{\mathsf{Fin}}}
\newcommand{\F}{\ensuremath{\mathbb{F}}}
\newcommand{\R}{\ensuremath{\mathcal{R}}\xspace}
\newcommand{\C}{\ensuremath{\mathcal{C}}\xspace}
\newcommand{\D}{\ensuremath{\mathcal{D}}\xspace}
\newcommand{\plus}{\ensuremath{\mathsf{plus}}}
\newcommand{\old}{\ensuremath{\mathsf{old}}}

\newcommand{\B}{\ensuremath{\mathcal{B}}\xspace}

\newcommand{\SSet}[1]{\ensuremath{{\mathsf{Supp}(#1)}}\xspace}
\newcommand{\NSet}{\ensuremath{{\mathsf{NSet}}}}
\newcommand{\Supp}[1]{\SSet{#1}}
\newcommand{\Idx}{\ensuremath{\mathbb{I}}}
\newcommand{\Indexed}{\ensuremath{{[|\Idx|,\Set]}}}
\newcommand{\Presheaf}{\ensuremath{{[\Idx,\Set]}}}
\newcommand{\EM}{\mathsf{EM}}
\newcommand{\colim}{\ensuremath{{\mathsf{colim}}}\xspace}

\newcommand{\textqt}[1]{`#1'}
\newcommand{\epito}{\ensuremath{\twoheadrightarrow}}
\newcommand{\monoto}{\ensuremath{\rightarrowtail}}
\newcommand{\hookto}{\ensuremath{\hookrightarrow}}
\renewcommand{\rho}{\varrho}

\newcommand{\dispind}{\qquad~}
\newcommand{\etal}[1][.]{et~al#1\xspace}
\newcommand{\N}{\ensuremath{\mathbb{N}}\xspace}

\newcommand{\Set}{\ensuremath{\mathsf{Set}}\xspace}
\newcommand{\Nom}{\ensuremath{\mathsf{Nom}}\xspace}
\newcommand{\RnNom}{\ensuremath{\mathsf{RnNom}}\xspace}
\newcommand{\OrdNom}{\ensuremath{\mathsf{OrdNom}}\xspace}
\newcommand{\restmon}[2][M]{\ensuremath{{#1/#2}}}
\newcommand{\id}{\ensuremath{\mathsf{id}}\xspace}
\newcommand{\Id}{\ensuremath{\mathsf{Id}}\xspace}
\newcommand{\pr}{\ensuremath{\mathsf{pr}}}
\newcommand{\fresh}{\ensuremath{\mathrel{\#}}}
\newcommand{\Abs}{\ensuremath{{[\A]}}}
\newcommand{\bind}{\ensuremath{{\lambda.}}}
\newcommand{\abs}[1]{\ensuremath{{\langle #1 \rangle}}}
\newcommand{\inl}{\ensuremath{\mathsf{in}_{1}}}
\newcommand{\inr}{\ensuremath{\mathsf{in}_{2}}}
\newcommand{\inj}{\ensuremath{\mathsf{in}}}

\newcommand{\set}[2][]{%
  \ifthenelse{\equal{#2}{}}{%
    \ensuremath{{#1\emptyset}}%
  }{%
    \ensuremath{{#1\{#2#1\}}}%
  }%
}

\newcommand{\Monad}[1][M]{\ensuremath{{#1\bullet}}}
\newcommand{\Pow}{\ensuremath{\mathcal{P}}}
\newcommand{\Powf}{\ensuremath{\mathcal{P}_{\mathsf{f}}}}
\newcommand{\Powufs}{\ensuremath{\mathcal{P}_{\mathsf{ufs}}}}

\newcommand{\ar}{\ensuremath{\mathsf{ar}}}
\newcommand{\supp}[1]{\ensuremath{\mathsf{supp}_{#1}}}
\newcommand{\maxidx}{\ensuremath{\text{\upshape maxidx}}}
\newcommand{\myold}{\ensuremath{\text{\upshape old}}}
\newcommand{\struct}{\ensuremath{\mathsf{struct}}}
\newcommand{\s}[1]{\ensuremath{\mathsf{s}_{#1}}}

\newcommand{\resteq}[1]{\ensuremath{\mathbin{\approx_{#1}}}}

\renewcommand{\hom}[1][]{\ensuremath{\mathsf{hom}_{#1}}}

\makeatletter
\ifthenelse{\boolean{proofsinappendix}}{
  \newoutputstream{proofstream}
  \openoutputfile{\jobname-proofs.out}{proofstream}
}{
}

\newcommand{\ExternalLink}{%
  \tikz[x=1.2ex, y=1.2ex, baseline=-0.05ex]{%
    \begin{scope}[x=1ex, y=1ex]
      \clip (-0.1,-0.1) 
      --++ (-0, 1.2) 
      --++ (0.6, 0) 
      --++ (0, -0.6) 
      --++ (0.6, 0) 
      --++ (0, -1);
      \path[draw, 
      line width = 0.5, 
      rounded corners=0.5] 
      (0,0) rectangle (1,1);
    \end{scope}
    \path[draw, line width = 0.5] (0.5, 0.5) 
    -- (1, 1);
    \path[draw, line width = 0.5] (0.6, 1) 
    -- (1, 1) -- (1, 0.6);
  }
}

\newcommand{\proofappendixbegin}[2]{%
  \phantomsection%
  \subsection*{\textbf{#1~\autoref{#2}}}%
  \addcontentsline{toc}{subsection}{#1~\autoref{#2}}%
  \label{#2:proof}%
  \def\proofappendix@qedsymbolmissing{\qed}
}
\newcommand{\proofappendixend}{%
  \proofappendix@qedsymbolmissing%
}
\let\oldqedhere\qedhere
\def\qedhere{\global\def\proofappendix@qedsymbolmissing{}\oldqedhere}

\newenvironment{proofhere}[2][Proof of]{%
  \subsection*{#1~\autoref{#2}.}%
  \def\proofappendix@qedsymbolmissing{\qed}%
}{%
  \proofappendixend%
  \par%
}

\ifthenelse{\boolean{proofsinappendix}}{%
  \NewEnviron{proofappendix}[2][Proof of]{%
    \ifthenelse{\boolean{linktoappendix}}{%
      \twskmarginpar{\hyperref[#2:proof]{\color{blue!40!black}{Appx.\ExternalLink}}}%
    }{}%
    \addtostream{proofstream}{\noexpand\proofappendixbegin{#1}{#2}}%
    \expandafter\addtostream{proofstream}{\expandonce{\BODY}}%
    \addtostream{proofstream}{\noexpand\proofappendixend}%
    \addtostream{proofstream}{}%
  }%
}{%
  \newenvironment{proofappendix}[2][Proof of]{%
    \begin{proofhere}[#1]{#2}
  }{%
    \end{proofhere}
  }
}
\makeatother

\addto\extrasenglish{

}
\addto\extrasUKenglish{

}

\tikzset{
  automaton/.style={
    initial text={},
    every edge/.append style={
      -stealth,
      thick,
      shorten <= 2pt,
      shorten >= 2pt,
      bend angle=20,
    },
    every label/.append style={
      align=center,
    },
  },
}
\usetikzlibrary{cd, calc, automata, fit, backgrounds, decorations, positioning}

\title{Supported Sets -- A New Foundation\texorpdfstring{\\}{} For Nominal Sets And Automata}
\titlerunning{Supported Sets}

\author{Thorsten Wißmann}{Radboud University, Nijmegen, the Netherlands\and\url{https://thorsten-wissmann.de}}{}{https://orcid.org/0000-0001-8993-6486}{Supported by the NWO TOP project 612.001.852}

\authorrunning{T.~Wißmann}

\Copyright{Thorsten Wißmann} %
\relatedversion{}
\relatedversiondetails{Full Version with Appendix}{http://arxiv.org/abs/2201.09825}



\category{} %

\useshorthands*{"}
\defineshorthand{"=}{\babelhyphen{hard}}

\acknowledgements{
  The author thanks Jurriaan Rot, Joshua Moerman, and Frits Vaandrager for
  inspiring discussions.
  The definition of supported sets originated from discussions with
  Lutz Schröder, Dexter Kozen, and Stefan Milius.
}

\begin{document}
\maketitle

\begin{abstract}
  The present work proposes and discusses the category of supported sets which provides a
  uniform foundation for nominal sets of various kinds, such as those for
  equality symmetry, for the order symmetry, and renaming sets. We show that all
  these differently flavoured categories of nominal sets are monadic over
  supported sets. Thus, supported sets provide a canonical finite way to
  represent nominal sets and the automata therein, e.g.~register automata. Name
  binding in supported sets is modelled by a functor following the idea of de
  Bruijn indices. This functor lifts to the well-known abstraction functor in
  nominal sets. Together with the monadicity result, this gives rise to a
  transformation process that takes the finite representation of a register
  automaton in supported sets and transforms it into its configuration automaton
  in nominal sets.
\end{abstract}

\section{Introduction}
Nominal sets provide an elegant framework to reason about structures that
involve names, the permutation of names, and name binding. Originally used
100 years ago by Fraenkel and Mostowski for the entirely different purpose of axiomatic
set theory, Gabbay and Pitts~\cite{GabbayP99} re-discovered them
in 1999 to define $\lambda$-expressions modulo $\alpha$-equivalence as an inductive data type.
Its associated structural recursion principle allowed it to define capture
avoiding substitution as a total function. Since then, a plethora of results
using nominal sets were established in many areas, including proof
assistants~\cite{UrbanT05,AydemirBW07}, calculi~\cite{GabbayC04,Staton07}, and
automata models~\cite{BojanczykEA14,skmw17,KozenMP015}. Nominal automata
are capable of processing words over infinite alphabets (\emph{data alphabets}),
while having good computational properties. Their expressiveness is similar (and
sometimes identical) to that of register
automata~\cite{KaminskiF94,DemriL09,CasselHJS16}, which are automata with a
\emph{finite} description processing infinite alphabets. This finiteness
condition translates into the notion of \emph{orbit\hyp{}finiteness} in the
nominal world, which requires extra work to obtain a finite description of
nominal automata, because orbit-finite objects are infinite in general.

In recent years, more general concepts of nominal sets were considered that
generalize from the permutation of names to other operations. One branch is called
\emph{renaming sets}~\cite{Staton07,GabbayH08,MoermanRot20} and allows that distinct names are
mapped to the same identifier; in another branch, symmetries on other data
alphabets~\cite{BojanczykEA14,VenhoekMR18} were considered, for example monotone
bijections on rational numbers $\Q$, called the \emph{total order symmetry}.

Nominal sets are not the only categorical approach to name binding. Multiple
presheaf-based approaches to name binding were developed over the
years~\cite{FiorePlotkinTuri99,FioreS06} which are strongly related
to nominal sets~\cite{GadducciEA2006}. Also, the theory of \emph{named sets}
were introduced for the study of history dependent automata~\cite{FMP02,MontanariP05}, they were shown to be equivalent (in the categorical sense) to nominal sets~\cite{GadducciEA2006}, and they also feature a name binding construction~\cite{CianciaM08}.
In contrast to nominal sets, the name binding in presheaf approaches and named
sets follow the method of \emph{de Bruijn indices}~\cite{debruijn1972}.

Despite their rich categorical structure, it is
known that the category of nominal sets (in all above\hyp{}mentioned flavours) is not
\emph{monadic} over sets, that is, their theory is not an algebraic theory that
can be described by a monad on sets -- unlike groups, rings, vector spaces, etc, which can be described by algebraic theories.

In order to still make algebraic methods applicable, we introduce
the category of \emph{supported sets} and show
that nominal sets are monadic over those. Thus, we make nominal sets
applicable to results and methods known from algebraic categories, such as the
generalized powerset construction~\cite{SilvaBBR13,bartelsphd} or results about
the representation of algebras~\cite{amsw19b} that provide a canonical finite
representation of orbit-finite nominal sets. Moreover, supported sets do not
require symmetries on the data alphabet, so they can also serve as a categorical
foundation
for data alphabets that are not described by symmetries.
In fact, we discovered supported sets while working on a learning algorithm for
register automata for general data alphabets.

\subparagraph*{Structure of the paper.} After recalling
the basic definitions for nominal sets (\autoref{secPrelim}), we introduce
supported sets and discuss their basic categorical properties
(\autoref{secSuppSet}). Having established the monadicity of nominal sets
(\autoref{secMonadicity}), we show that
supported sets have a functor for name binding that lifts to nominal sets
(\autoref{secAbstraction}). This yields a construction from (register) automata in supported sets
to nominal automata (\autoref{secGenDet}).
Full proofs and also additional explanations to definitions and examples can be found in the appendix.

\section{Preliminaries on Nominal Sets}
\label{secPrelim}
Before introducing supported sets in detail, we first recall some notations and basic concepts
of nominal sets. We assume that the reader is familiar with basic
category theory, but we restrict to set-theoretic definitions wherever
possible.

\begin{notation} \label{basicNotation}
  Given sets $X$, $Y$, the set of all maps $X\to Y$ is written $Y^X$.
Injective maps are denoted by
$\monoto$, surjective maps by $\epito$, and $\cdot$
denotes map composition. We fix sets $1 = \{0\}$, $2=\{0,1\}$.
The cycle notation
$(a_0 ~ a_1 \cdots a_{n-1})$
for elements of a set $A$
denotes the bijection $A\to A$ sending $a_0\!\mapsto\! a_1$, $a_1\!\mapsto\!a_2$, $\ldots$,
$a_{n-1}\!\mapsto\! a_0$, and fixing all other elements of~$A$.
\end{notation}
\begin{proofappendix}[Details for]{basicNotation}
  The precise definition of the cycle notation is as follows: Given elements
  $a_0$, \ldots $a_{n-1} \in A$, the notation
  \(
  (a_0 ~ a_1 \cdots a_{n-1})
  \)
  denotes the bijective map $f\colon A\to A$ given by
  \[
    f(x) = \begin{cases}
      a_{k+1\text{ mod }n} &\text{if there is some }k < n\text{ with }x = a_k
      \\
      x &\text{otherwise.}
    \end{cases}
    \tag*{\qed\qedhere}
  \]
\end{proofappendix}
Nominal sets and various flavours thereof are built around the notion of monoid
actions, which specifies how atoms nested in some structure can be permuted or
renamed.

\begin{definition}
  Given a monoid $(M,\cdot,e)$, an $M$-set is a set $X$ together with a map
  $\text{\upshape\textqt{\(\cdot\)}}\colon M\times X\to X$, called the \emph{action} and written
  infix $m\cdot x$ for $x\in X$, $m\in M$,
  such that $e\cdot x = x$ and $(m\cdot m')\cdot x = m\cdot (m'\cdot x)$ for all
  $m,m'\in M$, $x\in X$. Thus, we use {\upshape\textqt{$\cdot$}} both for the monoid multiplication
  and the action.
  A map $f\colon X\to Y$ between $M$-sets $(X,\cdot)$, $(Y,\star)$ is
  \emph{equivariant} if $f(m\cdot x) = m\star f(x)$ for all $m\in M$, $x\in X$.
\end{definition}

Throughout the paper, $M$ will be a submonoid of $(A^A,\cdot,\id_A)$ for a set
$A$, written $M\le A^A$, in which most of the results are parametric. The set
$A$ is called the set of \emph{atoms}, which can intuitively be understood as
names of registers or data-values that appear in the input of an automaton or as
names used for binding operations. The submonoid $M$ determines a subset of maps
of interest, closed under composition. Thus,
we use $\cdot$ both for map composition and monoid-multiplication, and moreover,
the unit of the monoid $M$ is simply $\id_A$.

\begin{example}
  \label{MonList}
  The main instances of monoids $M$ for nominal techniques are as follows:
\begin{enumerate}[topsep=0pt,beginpenalty=99,midpenalty=0]
\item For a set $A$, let $\Perm(A)$
  be the bijections on $A$ that modify only finitely many elements, i.e.
  \[
    \Perm(A) := \big\{ \phi\colon A\to
    A\mid \phi \text{ bijective and }\{a\in A\mid \phi(a) \neq a\}\text{ is finite}\big\}
  \]
  For nominal sets (over the equality symmetry), one
  fixes a countably infinite set $\A$ (understood as \emph{names}) and fixes $M := \Perm(\A)$~\cite{GabbayP99,pittsbook}.

\item For the set \Q of rational numbers, let $\Aut(\Q,<)$ be the set of
  bijective and monotone maps $\Q\to \Q$. For nominal sets over the total order
  symmetry, one considers $M := \Aut(\Q,<)$~\cite{BojanczykEA14}.

\item Let $\Fin(A)\subseteq A^A$ be the set of maps $f\colon A\to A$ for
  which $\{f(a)\neq a\mid a \in A\}$ is finite. For \emph{nominal renaming
    sets}, one considers $A :=\A$ and $M:= \Fin(\A)$~\cite{GabbayH08}.
\end{enumerate}
\end{example}

\noindent
An element $x\in X$ of an $M$-set $(X,\cdot)$ can be understood as a
structure with elements from $A$ embedded.
We can alter these embedded elements according to $m\in M$, yielding $m\cdot
x\in X$.
\begin{example} \label{mSetEx}
  For every set $A$ and $M\le A^A$, the following sets are $M$-sets:
  \begin{enumerate}
  \item The set $A$ itself with the action $m\cdot a := m(a)$, for $m\in M$,
    $a\in A$.
  \item\label{exPowf} If $X$ is an $M$-set, then so is $\Powf(X)$, the set of finite subsets
    of $X$, where the action is defined point-wise:
    \(
      m\cdot S := \{m\cdot x\mid x\in S\}
      ~\text{for }m\in M, S\in\Powf X.
    \)
  \item\label{exDiscrete} Every set $D$ can be equipped with the \emph{discrete} action:
    \(
      m\cdot d = d
      \text{ for all }m\in M, d\in D.
    \)
  \item\label{mLeft} The set $M$ itself is an $M$-set with monoid multiplication
    $M\times M\to M$ as the action.
  \end{enumerate}
\end{example}
\begin{proofappendix}[Details for]{mSetEx}
  All examples listed here satisfy the $M$-set axioms:
  \begin{enumerate}
    \item[1.] $A$ is an $M$-set because indeed $\id_A\cdot a = \id_A(a) = a$
      and $(\ell\cdot m)\cdot a = \ell(m(a)) = \ell\cdot (m\cdot a)$.
    \item[\ref{exPowf}.] \& \ref{exDiscrete}. The verification of the $M$-set axioms for $\Powf(X)$ and discrete
      $M$-sets are straightforward.
    \item[\ref{mLeft}.]
      For disambiguation, let us use $\circ$ for monoid multiplication
      $(M,\circ,e)$ and
      $\star$ for the monoid action, i.e.~for $m,x\in M$, we define
      \[
        m\star x := m\circ x
      \]
      Then, the monoid laws directly imply that $\star$ satisfies axioms of the $M$-set action:
      \[
        e\star x = e\circ x = x
      \]
      \[
        m'\star (m\star x)
        = m'\circ (m\circ x)
        = (m'\circ m)\circ x
        = (m'\circ m)\star x
        \tag*{\qedhere}
      \]
  \end{enumerate}
\end{proofappendix}

\noindent
The outcome of $m\cdot x$ only depends on what $m$ does on
the atoms $S\subseteq A$ buried in $x$:
\begin{definition}
  \label{defRestEq}
  We say that $m,m'\in M$ are \emph{identical on $S\subseteq A$}, written
  $m\resteq{S}m'$, if $m(a) = m'(a)$ for all $a\in S$.
\end{definition}

With the intuition that the action of an $M$-set $X$ renames the atoms $A$
in an element $x\in X$, we can derive from the action which atoms
an element $x\in X$ carries. Then, an $M$-set is \emph{nominal}, if each $x\in
X$ carries only finitely many atoms.
\begin{notheorembrackets}
\begin{definition}[{\cite[Def.~7]{GabbayH08}}]
  \label{defNominal}
  A set $S\subseteq A$ \emph{supports} $x\in X$ of an $M$-set $X$, if
  for all $m\resteq{S} m'$ (in $M$), we have $m\cdot x = m'\cdot x$.
  An $M$-set $X$ is \emph{nominal} if every element of $X$ has some finite support,
  i.e.~is supported by a
  finite set $S\subseteq A$.
\end{definition}
\end{notheorembrackets}
\begin{proofappendix}[Details for]{defNominal}
  Given that $M$ is a group, $S\subseteq A$ supports $x\in X$ iff
  for all $m\in M$ with
  \(\forall a\in S\colon m(a) = a\),
  we have $m\cdot x = x$.

  This is indeed equivalent to \autoref{defNominal}:
  For \textqt{only if}, put $m' := \id_A$. For \textqt{if}, take $m,m'\in M$
  with $m(a) = m'(a)$ for all $a\in S$. Hence, $(m'^{-1}\cdot m)(a) = a$ for all
  $a\in S$, hence $m\cdot x = m'\cdot m'^{-1}\cdot m\cdot x = m'\cdot x$.

  The definition can be simplified in the same way for nominal renaming sets~\cite[Lem.~13]{GabbayH08}.
\end{proofappendix}
\noindent
If $M$ happens to be a group, then the definition of support can be simplified slightly~\cite{GabbayP99,pittsbook}.

\begin{example} \label{nomSetList}
  Most of the $M$-sets from \autoref{mSetEx} are nominal:
  \begin{enumerate}[topsep=0pt]
  \item $A$ is nominal: every $a\in A$ is supported by $S := \set{a}$,
    and also by any superset of $S$.
  \item If $X$ is a nominal $M$-set, then so is $\Powf(X)$. A set $E\in\Powf(X)$
    of elements is supported by the union of finite supports of the elements $x\in E$.
    This union is finite because $E$ is so.
  \item Every discrete $M$-set $D$ is nominal because every $x\in D$ is
    supported by the empty set. 
  \item\label{mLeftNotNom} For all monoids $M$ of interest for nominal
    techniques, $M$ considered as an $M$-set is not
    nominal: whenever $M$ is a nominal $M$-set, then \emph{every} $M$-set is
    nominal.
    For example, $\Perm(\A)$ is not nominal because no $\sigma \in
    \Perm(\A)$ has finite support.
  \end{enumerate}
\end{example}
\begin{proofappendix}[Details for]{nomSetList}
  The examples of nominal $M$-sets \autoref{nomSetList} are all standard. It
  only remains to verify the non-example in item \ref{mLeftNotNom}, i.e.~let us
  verify that whenever $M$ is a nominal $M$-set, then all $M$-sets are nominal.

  If $M$ is a nominal $M$-set, let $S\subseteq A$ be a finite set that supports
  $\id_A\in M$. Then, $S$ is the support of all elements $x\in X$ in all
  $M$-sets $X$: consider $m,m'\in M$ with $m\resteq{S} m'$. Then
  $m= m\cdot \id_A = m'\cdot \id_A = m'$ because $S$ supports $\id_A$. Hence,
  $m\cdot x = m'\cdot x$, showing that $S$ also supports $x\in X$.
\end{proofappendix}

\begin{definition}
  Let $\Nom(M)$ be the category of nominal $M$-sets and equivariant maps.
\end{definition}

\begin{enumerate}
\item The category of \emph{nominal sets}
  is denoted by $\Nom = \Nom(\Perm(\A))$ (slightly overloading notation), that
  is for $A:=\A$ and $M:=\Perm(\A)$. This is called \emph{the 
  equality symmetry}.
\item \emph{Ordered nominal sets} are $\OrdNom=\Nom(\Aut(\Q,<))$,
  i.e.~for $A:=\Q$, $M:=\Aut(\Q,<)$.
\item \emph{Nominal renaming sets} are $\RnNom$ $=$ 
  $\Nom(\Fin(\A))$, i.e.~for $A:=\A$, $M:=\Fin(\A)$.
\end{enumerate}

\begin{notheorembrackets}
\begin{definition}[{\cite[Definition 4.11]{BojanczykEA14}}]
  A monoid $M\le A^A$ \emph{admits least supports} if each element of a nominal
  $M$-set has a least finite support.
  If so, we write $\supp{X}\colon X \to \Powf(A)$ for the map that sends elements of
  the $M$-set $X$ to their least finite support.
\end{definition}
\end{notheorembrackets}
In all running examples of nominal set flavours, the monoid admits least
supports,
i.e.~$\Nom$~\cite{GabbayP02,pittsbook}, $\OrdNom$~\cite{BojanczykEA14}
and $\RnNom$~\cite{GabbayH08}. Intuitively, $\supp{}(x)\subseteq A$ can be
understood as precisely the atoms that appear in
$x$. For example, $\supp{A}\colon A\to \Powf(A)$ sends $a$ to
$\{a\}$, and $\supp{\Powf(X)}\colon \Powf(X)\to \Powf(A)$ is
\(
  \supp{\Powf(X)}(E) = \smash{\bigcup}\{\supp{X}(x)\mid x\in E\}.
\)
The opposite of an atom $a\in A$ in the support is:
\begin{definition} An atom $a\in A$ is \emph{fresh} for $x\in
  X$ in a nominal set $X$ (notated
  $a\fresh x$), if $a\notin
  \supp{X}(x)$. For multiple elements, we write $a\fresh x,y$ to denote that
  $a$ is fresh for $x$ and $y$.
\end{definition}

\noindent
Both freshness and support are preserved by equivariant maps.
Intuitively, equivariant maps can possibly forget about atoms, but can never
introduce new atoms:
\begin{lemma}
  \label{equivSupport}
  For an equivariant map, if $S$ supports $x$,
  then $S$ also supports $f(x)$. If $X$ and $Y$ have least finite supports, then
  $\supp{Y}(f(x))\subseteq \supp{X}(x)$.
\end{lemma}
\begin{proofappendix}{equivSupport}
  If $S$ supports $x$, consider $m,m'\in M$ with $m\resteq{S}m'$. Then,
  $m\cdot f(x) = f(m\cdot x) = f(m' \cdot x) = m'\cdot f(x)$.

  For least finite supports, we have that $\supp{X}(x)$ supports $f(x)$ by the
  previous statement. Hence, the least finite support $\supp{Y}(f(x))$ is
  smaller or equal to $\supp{X}(x)$ (w.r.t.~subset inclusion).
\end{proofappendix}
\noindent The set of elements in an $M$-set that can be reached from an element $x$ is
called the orbit:
\begin{definition}
  Given a group $M$, the \emph{orbit} of an element $x$ in an $M$-set is the
  subset $\{m\cdot x\mid m\in M\}\subseteq X$. A nominal $M$-set is
  \emph{orbit-finite} if it consists of finitely many orbits.
\end{definition}
In this definition, we assume $M$ to be a group, because then,
every nominal set $X$ is a disjoint union of orbits. For $M:=\Perm(\A)$, the
\emph{orbit-finite} nominal sets are precisely the finitely presentable objects in
$\Nom$~\cite[Proposition~2.3.7]{petrisanphd}. E.g.~$\A$ is orbit-finite,
$\Powf\A$ is not. In nominal automata, one requires the state set to be orbit-finite, as opposed to finiteness in classical automata theory.

\section{Supported Sets}
\label{secSuppSet}
The central notion of the present paper are supported sets, which are parametric
in the set $A$ of atoms or data symbols of interest:

\begin{definition}
  For a set $A$, the category of \emph{supported sets} $\SSet{A}$ contains the
  following:%
\\
  \begin{minipage}{.8\textwidth}
  \begin{enumerate}
  \item a supported set $X$ is a set $X$ together with a map
    $\s{X}\colon X\to \Powf(A)$.
  \item a supported map $f\colon (X,\s{X}) \to (Y,\s{Y})$ is a map $f\colon X\to Y$
    with $\s{Y}(f(x))\subseteq \s{X}(x)$ for all $x\in X$. This means, $f$ makes
    the right-hand triangle weakly commute.
  \end{enumerate}
\end{minipage}%
\begin{minipage}{.2\textwidth}
  \hfill
  \(
    \begin{tikzcd}[column sep=0pt]
      X
      \arrow{rr}{f}
      \arrow{dr}[swap]{\s{X}}
      &
      {} \descto[pos=.3]{d}{\(\supseteq\)}
      & Y
      \arrow{dl}{\s{Y}}
      \\
      & \Powf(A)
    \end{tikzcd}
  \)
\end{minipage}

  Whenever clear from the context, we simply speak of supported sets $X, Y\in
  \SSet{A}$ and supported maps $f\colon X\to Y$, leaving $\s{X}$ and $\s{Y}$ implicit.
\end{definition}

The intuition for supported sets comes from the least finite support map
$\supp{}$ of nominal sets. Hence, in a supported set $X$, the map $\s{X}(x)$ tells
which atoms are carried by $x\in X$, but we can not permute or rename them in
general.

The definition of the morphisms in supported sets comes from the property
of equivariant maps: they possibly forget about atoms in the support, but
they can never introduce new atoms (\autoref{equivSupport}). This observation
becomes the defining property of supported maps.

\begin{example}
  The set $A$ itself is a support set with $\s{A}(a) = \{a\}$. However, the
  singleton subset $\{a\}\subseteq A$ is also a supported set with
  $\s{\set{a}}(a) = \set{a}$.
  In general, every nominal $M$-set $X$ (for $M$ admitting least supports) yields
  a supported set by putting $\s{X} := \supp{X}$.
\end{example}

The difference between nominal and supported sets is that $\supp{}$ in a nominal
set $X$ is a derived notion since it is implicit in the $M$-set action. On the
other hand, for supported sets, $\s{X}$ is part of the syntactical structure.
For the sake of clarity, we use different mathematical symbols for the
semantical $\supp{}$ and the structural $\s{}$.

\begin{lemma} \label{forgetful}
  For every $M\le A^A$,
  there is a faithful functor $U\colon \Nom(M)\to
  \SSet{A}$ sending $(X,\cdot)$ to a supported set $(X,\s{X})$, where
  $\s{X}(x)$ is the intersection of all finite supports of $x$ in $(X,\cdot)$.
  If $M\le A^A$ admits least supports, then $\s{UX} = \supp{X}$.
  Equivariant maps in $\Nom(M)$ are sent to their underlying map by $U$.
\end{lemma}

\begin{proofappendix}{forgetful}
  For arbitrary $M\le A^A$, define the functor $U\colon \Nom(M)\to \SSet{A}$
  sending $(X,\cdot)$ to the set $X$ with the support
  \[
    \s{X}(x) = \bigcap\{S\subseteq A\mid S\text{ finite and supports }x \}
  \]

  Since every element $x\in X$ has some finite support (\autoref{defNominal}), the intersection
  in the definition of $\s{X}$ is not empty, and hence $\s{X}(x)$ is a
  finite set and so $U$ sends nominal $M$-sets to supported sets.

  For an equivariant map $f\colon X\to Y$ and $x\in X$, we have
  by \autoref{equivSupport}:
  \begin{align*}
    &\{S\subseteq A\mid S\text{ finite and supports }f(x) \}
    \\ &
    \qquad\supseteq \{S\subseteq A\mid S\text{ finite and supports }x \}.
  \end{align*}
  When taking the intersection of these families of
  subsets of $A$, the inclusion reverses and we obtain
  \begin{align*}
    \s{Y}(f(x))
    &=\bigcap\{S\subseteq A\mid S\text{ finite and supports }f(x) \}
      \\ &
    \subseteq \bigcap\{S\subseteq A\mid S\text{ finite and supports }x \}
    = \s{X}(x)
  \end{align*}
  Thus, $Uf$ is a supported map. Clearly, $U$ preserves identities and
  composition. Note that we do did not require here that \emph{least} finite
  supports exist in nominal $M$-sets.

  If they exist however, we have that the least finite support $\supp{X}(x)$ of
  $x\in X$ is the intersection of all finite supports of $x$, i.e.~$\s{X}(x)
  = \supp{X}(x)$, as claimed.
\end{proofappendix}

Later, we show that this forgetful functor $\Nom(M)\to\Supp{A}$ is
right-adjoint and even monadic, so one can consider nominal sets as algebras on
supported sets. Before, we establish some categorical properties of $\Supp{A}$
(generic in the choice of $A$) that will come in handy.

\subsection{Universal Constructions and Finiteness}
\label{suppCat}
Unsurprisingly, supported sets have a very set-like nature.
In the present \autoref{suppCat}, we let $\ToSet$ denote the forgetful functor
$\ToSet\colon \Supp{A}\to \Set$ that forgets the support map and sends $(X,\s{X})$ to
the plain set $X$. Conversely, every set can be equipped with trivial support:
\begin{lemma} \label{setSubcat} The inclusion functor $J\colon
  \Set\hookto\Supp{A}$
  defined by
  \(
  JX=(X,\s{X})
  \)
  with $\s{X}(x)=\emptyset$
  is right-adjoint to the forgetful
  $\ToSet\colon \Supp{A}\to \Set$ ($\ToSet \dashv J$) and $\ToSet J=\Id_{\Set}$.
\end{lemma}
\begin{proofappendix}{setSubcat}
  Define $J\colon \Set\hookto \Supp{A}$ by $J X = (X,\s{X})$, $\s{X}(x) =
  \emptyset$. Every map $f\colon X\to Y$ then is a supported map $f\colon JX\to
  JY$.
  Clearly, $\ToSet JX=X$ and $\ToSet J(f\colon X\to Y) = f$, hence $\ToSet J=\Id_{\Set}$.

  For the adjunction,
  we verify that there are supported maps $\eta_X\colon X\to J\ToSet X$ such that
  for every supported map $h\colon X\to JY$ ($Y\in \Set$),
  there is a unique $h'\colon \ToSet X\to Y$ with $h = Jh'\cdot \eta_X$:
  \[
    \begin{tikzcd}
      J\ToSet X
      \arrow{r}{Jh'}
      & JY
      \\
      X
      \arrow{u}{\eta_X}
      \arrow{ur}[swap]{h}
    \end{tikzcd}
  \]
  Note that the identity $\id_X\colon X\to X$ gives rise to the supported map
  $\eta_X\colon X\to J\ToSet X$. Hence, $Jh'\cdot \eta_X = h$ and that $h' := \ToSet h$.
  Since $\eta_X$ is surjective and $J$ is faithful, $h'$ is the unique map
  fulfilling this property.
\end{proofappendix}
In other words, $\Set$ is a reflective subcategory of $\Supp{A}$.
Similarly to sets, universal constructions such as limits and colimits all exist
in supported sets and are (almost) set-like. Given a diagram $D\colon
\D\to\Set$, we write $\pr_X\colon \lim D\to DX$ for the limit projection map of
$X\in \D$ and $\inj_X\colon DX\to \colim D$ for the colimit injections.

\begin{proposition}\label{suppColim}
  All colimits in $\Supp{A}$ exist:
  Given a diagram $D\colon \D\to \Supp{A}$, the colimit is formed in \Set and then
  equipped with the support $\s{}\colon
  \colim\, \ToSet D\to \Powf(A)$:
  \[
    \s{}(c) = \bigcap \{ \s{DY}(y)\mid Y\in \D, y\in DY, \inj_Y(y) = c \}.
  \]
  \twnote{}
\end{proposition}
\begin{proofappendix}{suppColim}
  Consider a diagram $D\colon \D\to \Supp{A}$ and its colimiting cocone in \Set:
  \[
    \inj_{Y}\colon DY\to C
    \qquad\text{in }\Set.
  \]
  In order to make $C$ a supported set, define $\s{}\colon C\to \Powf(A)$ by
  \[
    \s{}(c) = \bigcap \{ \s{DY}(y)\mid Y\in \D, y\in DY, \inj_Y(y) = c \}
  \]
  Since colimit injections in $\Set$ are jointly surjective, the above
  intersection is non-empty, and thus yields a finite subset of~$A$.
  
  This lets the cocone lift to $\Supp{A}$: every $\inj_Z\colon DZ\to C$ is a
  supported map, because for every $z\in DZ$ we have
  \begin{align*}
    \s{C}(\inj_Z(z))
    &= \bigcap \{ \s{DY}(y)\mid Y\in \D, y\in DY, \inj_Y(y) = \inj_Z \}
    \\ &
    \subseteq \s{DZ}(z).
  \end{align*}
  In order to see that $(\inj_Y)_{Y\in \D}$ is a colimiting cocone, consider
  a cocone $(e_Y\colon DY\to E)_{Y\in \D}$ in $\Supp{A}$. Since this is also a
  competing cocone in $\Set$, we obtain a cocone morphism $u\colon C\to E$.
  For the verification that $u$ is a supported map, note that for every
  $c\in C$ and every $Y\in \D$, $y\in DY$ with $\inj_Y(y) = c$,
  we have $\s{E}(u(c)) \subseteq \s{Y}(y)$. By the universal property of intersection,
  this implies that
  \[
    \s{E}(u(c)) \subseteq \bigcap\{\s{Y}(y)\mid Y\in \D, y\in DY, \inj_Y(y) = c\}
    = \s{C}(c),
  \]
  so $u\colon C\to E$ is a supported map. Since $\ToSet \colon \Supp{A}\to \Set$ is
  faithful, $u$ is a cocone morphism in $\Supp{A}$ and moreover unique.
\end{proofappendix}
The intersection handles the case where elements of possibly
different support are identified in the colimit. Thus, the intersection vanishes if there are no morphisms in the diagram:

\begin{example} \label{suppCoprod}
  The coproduct of supported sets $X$, $Y$ is given by their disjoint
  union, $X+Y$ equipped with support $\s{X+Y}(\inj_X(x))
  = \s{X}(x)$ and $\s{X+Y}(\inj_Y(y))= \s{Y}(y)$.
\end{example}
\begin{proofappendix}[Verification of]{suppCoprod}
  We show the statement about coproducts directly for $I$-indexed families
  $(X_i)_{i\in I}$:
  by \autoref{suppColim}, the coproduct $\coprod_{i\in I} X_i$ in $\Supp{A}$ is given by the coproduct
  in \Set. For every $c\in \coprod_{i\in I}X_i$ there is precisely one $i\in I$,
  $x\in X_i$ with $\inj_i(x) = c$. Hence, $\s{\coprod_{i\in I}X_i}(c) = \s{X_i}(x)$.
\end{proofappendix}
\noindent
Limits on the other hand are formed differently because of the finite support map:

\begin{proposition} \label{suppLim} $\Supp{A}$ is complete. The limit of a
  diagram $D\colon \D\to\Supp{A}$ is the subset of finitely supported elements
  in the limit $\lim \ToSet D$ in \Set:
  \begin{equation}
    \lim D = \{
        x\in \lim \ToSet D
        \mid \bigcup_{Y\in \D} \s{DY}(\pr_Y(x))\text{ is finite}
      \}
      \label{supportOnLimit}
  \end{equation}
\end{proposition}
\begin{proofappendix}{suppLim}
  Consider a diagram $D\colon \D\to \Supp{A}$ and its limiting cone in \Set:
  \[
    \pr_{Y}\colon P\to DY \qquad\text{in }\Set.
  \]
  Define the map to all -- i.e.~possibly infinite -- subsets of $A$:
  \[
    p\colon P\to \Pow(A)
    \qquad
    p(x) = \bigcup\{\s{DY}(\pr_{Y}(x)) \mid Y\in \D \}
  \]
  The limit in $\Supp{A}$ is the restriction of $P$ to finitely supported
  elements:
  \[
    L := \{ x\in P\mid p(x)\text{ is finite}\}
    \qquad \s{L}(x) := p(x).
  \]
  By definition, $L$ is a finitely supported map. The projections are the
  restrictions of $\pr_Y$ to $L\subseteq P$: $\ell_Y\colon L\to DY$ and $\ell(x)
  = \pr_Y(x)$ for all $Y\in \D$. Hence,
  \[
    \s{L}(x) = \bigcup\{
    \s{DY}(\ell_Y(x))\mid Y\in \D
    \}.
  \]
  It is immediate that every $\ell_Y$ is a
  supported map:
  \[
    \s{DY}(\ell_Y(x)) \subseteq
    \bigcup\{\s{Z}(\ell_Z(x)) \mid
    Z\in \D
    \}
    = \s{L}(x)
  \]
  The family $(\ell_Y\colon L\to DY)_{Y\in \D}$ is a cone because it was defined
  as a restriction of the cone $P$ in \Set. For the verification of the
  universal property, consider another cone
  \[
    (e_Y\colon E\to DY)_{Y\in \D}
    \qquad\text{in }\Supp{A}.
  \]
  This is also a cone for the diagram $\ToSet D\colon \D\to \Set$, so we obtain a
  unique cone morphism $u \colon E\to P$ (i.e.~$\pr_Y\cdot u=e_Y$ for all $Y\in \D$). 
  For every $x\in E$ and $Y\in \D$, we have
  \[
    \s{DY}(\pr_Y(u(x))) = \s{DY}(e_Y(x)) \subseteq \s{E}(x)
  \]
  since $e_Y$ is a supported map. So by the universal property of union, we obtain
  \[
    p(u(x)) = \bigcup\{\s{DY}(\pr_Y(u(x)))\mid Y\in \D\} \subseteq \s{E}(x),
  \]
  which proves the finiteness of $p(u(x))\subseteq A$. Hence, $u\colon E\to P$
  restricts to $u\colon E\to L$ which is thus also a supported map ($\s{L}(u(x))
  = p(u(x)) \subseteq s_E(x)$). Uniqueness of $u\colon U\to L$ follows directly
  from the faithfulness of $\ToSet\colon \Supp{A}\to \Set$. Hence, $(\ell_Y\colon
  L\to DY)_{Y\in \D}$ is a limiting cone.
\end{proofappendix}
\noindent
The process of restricting to finitely supported elements also happens for limits
in nominal sets. For finite limits, this side-condition disappears:
\begin{corollary} \label{suppFinLim}
  $\ToSet \colon \Supp{A}\!\to\!\Set$ preserves all finite limits.
\end{corollary}
\begin{proofappendix}{suppFinLim}
  Consider a diagram $D\colon \D\to \Supp{A}$ and its limit
  \[
    (\pr_Y\colon \lim \ToSet D \to \ToSet DY)_{Y\in \D}
    \quad\text{ in $\Set$}.
  \]
  If $\D$ has only finitely many objects, then
  \[
    \bigcup\{\s{DY}(\pr_Y(x)) \mid Y\in \D\}
  \]
  is finite for all $x\in \lim \ToSet D$. Hence, the limit in $\Supp{A}$ given by
  \autoref{suppLim} is identical to the limit $\lim \ToSet D$ in $\Set$.
\end{proofappendix}

$\Supp{A}$ is cartesian closed, that is, we have an internal hom object.
Similarly to nominal set, this internal hom object $X^E$ in $\Supp{A}$
contains more than just the hom set $\hom(E,X)$:

\begin{definition} \label{exponent}
  For supported sets $X$ and $E$, let $X^E$ contain
  those maps $f\colon E\to X$
  for which $\s{X^E}(f) = \bigcup_{e\in E}\s{X}(f(e))\setminus \s{E}(e)$ is finite.
\end{definition}
\begin{proofappendix}[Details for]{exponent}
  For a supported set $E$, the full definition of the functor $(-)^E\colon
  \Supp{A}\to\Supp{A}$ is
  \[
    X^E := \{f\colon E\to X\mid \bigcup_{e\in E}\s{X}(f(e))\setminus
    \s{E}(e)\text{ is finite}\}
    \quad
    \s{X^E}(f) = \bigcup_{e\in E}\s{X}(f(e))\setminus \s{E}(e)
  \]
  on supported sets $X$. For supported maps $g\colon X\to Y$ we put
  \(
  g^E\colon X^E\to Y^E,
  g^E(f) = g\cdot f.
  \)
\end{proofappendix}
Note that \textqt{\emph{map}} really means ordinary maps between sets.
Intuitively, a map $f\in X^E$ may introduce finitely many atoms when mapping
elements of $E$ to $X$.
For example, if $E$ is a finite supported set, then $X^E$ contains all maps
  $E\to X$.
In contrast, a \emph{supported} map may not introduce any new atoms at all,
hence, a map $f\colon E\to X$ is a supported map if and only if $\s{X^E}(f) =
\emptyset$. In particular, $\hom(E,X)\subseteq X^E$.

\begin{proposition}
  \label{suppCCC}
  $(-)\times E$ is left adjoint to $(-)^E$,
  for all supported sets $E$.
\end{proposition}
\begin{proofappendix}{suppCCC}
  Define $\eta_X\colon X\to (X\times E)^E$ by
  \[
    \eta_X(x)(e) = (x,e).
  \]
  For every $x\in X$, clearly $\eta_X(x)$ is in $(X\times E)^E$ since
  \[
    \s{(X\times E)^E}(\eta_X(x))
    = \bigcup_{e\in E} \s{X\times E}(x,e) \setminus \s{E}(e)
    = \bigcup_{e\in E} \s{X}(x) = \s{X}(x)
  \]
  is finite; this also shows that $\eta_X$ is a supported map.
  Moreover, $\eta_X$ is natural in $X$ since
  \begin{align*}
    (g\times E)^E(\eta_X(x) &= (g\times E)\cdot (e\mapsto (x,e))
                              \\ &
    = (e\mapsto (g(x), e) = \eta_Y(g(x)).
  \end{align*}
  For the verification of the universal mapping property, consider $g\colon X\to
  Y^E$ and define
  \[
    h\colon X\times E\to Y
    \quad
    \text{ by }
    \quad
    h(x,e) = g(x)(e).
  \]
  By the definition of $\s{Y^E}$, we have
  \[
    \s{Y}(g(x)(e)) \setminus \s{E}(e) ~~\subseteq~~ \s{Y^E}(g(x))
  \]
  and so
  \[
    \s{Y}(g(x)(e)) \subseteq \s{Y^E}(g(x))\cup\s{E}(e) 
    \subseteq \s{X}(x)\cup\s{E}(e) 
    = \s{X\times E}(x,e),
  \]
  hence, $h$ is a supported map. For all $x\in X$, we have
  \begin{align*}
    h^E(\eta_X(x)) &= h\cdot (\eta(x)) = h\cdot (e\mapsto (x,e))
    \\ &
    = (e\mapsto h(x,e))
    = (e\mapsto g(x)(e))
    = g(x),
  \end{align*}
  so $h^E\cdot \eta_X = g$. For uniqueness, consider a supported map $u\colon
  X\times E\to Y$ with $u^E\cdot \eta_X = g$. Necessarily,
  \[
    g(x) = u^E(\eta_X(x)) = u\cdot (e\mapsto (x,e)) = (e\mapsto u(x,e))
  \]
  i.e.~$u(x,e) = g(x)(e) = h(x,e)$ and $u=h$.
\end{proofappendix}

In contrast to nominal sets, categorical finiteness notions in $\Supp{A}$ express
actual finiteness. For the present paper, we do not need the precise
definition of finitely presentable objects and locally finitely presentable (lfp)
categories~\cite{adamek1994locally,gu71}. For the present purposes, it is enough to mention that $\Supp{A}$
is lfp and that the finitely presentable objects are the finite
supported sets, i.e.~$\Supp{A}$ is locally finite. Detailed definitions can be found in the proof.
\begin{proposition} \label{suppLFP}
  $\Supp{A}$ is lfp and finite presentability is actual finiteness.
\end{proposition}
\begin{proofappendix}{suppLFP}
\begin{notheorembrackets}
  Let us first recall the definition of fp objects and lfp categories:
\begin{definition}[\cite{adamek1994locally,gu71}]
  \begin{enumerate}[topsep=0pt]
  \item A \emph{directed colimit} is the colimit for a diagram $D\colon \D\to \C$
    where $\D$ is a poset in which any two objects have an upper bound.
  \item An object
  $X$ in a category $\C$ is called \emph{finitely presented (fp)} if the hom-functor
  $\C(X,-)\colon \C\to \Set$ preserves colimits of directed diagrams.
  \item A category $\C$ is called \emph{locally finitely presentable (lfp)} if it is cocomplete and every object is
  the directed colimit of finitely presentable objects.
\end{enumerate}
\end{definition}
\end{notheorembrackets}

Given $X\in \Supp{A}$ and a directed diagram $D\colon \D\to
\Supp{A}$, note that the colimit preservation
\[
  \Supp{A}(X, \colim D)
  = \colim \Supp{A}(X, D(-))
\]
boils down to the following condition~\cite[Def.~1.1]{adamek1994locally}:
for every supported map $f\colon X\to \colim D$, there exists $Y\in \D$ such that
\begin{enumerate}
\item there are $Y\in \D$ and $g\colon X\to DY$ with $\inj_Y\cdot f' = f$:
  \[
    \begin{tikzcd}
      X
      \arrow{r}{f}
      \arrow{dr}[swap]{g}
      & \colim D
      \\
      & DY
      \arrow{u}[swap]{\inj_Y}
    \end{tikzcd}
    \qquad\text{in }\Supp{A}
  \]
\item $Y$ and $g$ are essentially unique in the sense that if $f =
  \inj_{Y'}\cdot g'$ for $g'\colon X\to DY'$, then $D(Y\to Z)\cdot g = D(Y'\to
  Z)\cdot g'$ for some $Z\ge Y$ in $\D$ (recall that $\D$ is a poset, so we can
  denote connecting morphisms simply by $Y\to Z$ and $Y'\to Z$)
\end{enumerate}

We now prove the following statements:

\begin{enumerate}
\item\label{itemDirColimSubset}
  Every supported set $X$ is the directed colimit of its finite subsets:

  Let $\D = \{ Y\mid Y \subseteq X\}$, $D\colon \D\to \Supp{A}$, $DY = (Y,\s{DY})$ where $\s{DY}$ is the restriction of
  $\s{X}\colon X\to \Powf(A)$ to $Y\subseteq X$. In $\D$, any two elements
  $Y,Y'\in \D$ have an upper bound: $Y\cup Y'\in \D$. By \autoref{suppColim}, we have
  that $\colim D$ is carried by $X$, and
  \begin{align*}
    \s{\colim D}(x)
    &= \bigcap \{ \s{DY}(x) \mid Y\in \D, x\in Y\}
     \\
    &= \bigcap \{ \s{X}(x) \mid Y\in \D, x\in Y\}
    = \s{X}(x).
  \end{align*}
  So $\colim D \cong X$, as desired.

\item Every finitely presentable supported set $X$ is finite:

  Consider the directed colimit of subsets of $X$. By
  \autoref{itemDirColimSubset}, this directed colimit yields $X$, so we have an
  isomorphism $\phi\colon X\to \colim D$. Since $X$ is finitely presentable,
  this factors through some finite subset $Y\subseteq X$:
  \[
    \begin{tikzcd}
      X
      \arrow{r}{\phi}[swap]{\cong}
      \arrow[bend right]{dr}[swap]{g}
      & \colim D
      \\
      & DY
      \arrow{u}[swap]{\inj_Y}
    \end{tikzcd}
    \qquad\text{in }\Supp{A}
  \]
  Note that $\inj_Y$ is support-preserving by the definition of
  $\D$ and injective since it's the same colimit injection as in \Set. Also,
  $\inj_Y$ is surjective, since $\inj_Y\cdot g = \phi$. Hence, it is a regular
  monomorphism, an epimorphism and thus an isomorphism, showing finiteness of $X$.

\item Every finite supported set $X$ is finitely presentable:

  Since finitely presentable objects are closed under finite
  coproducts~\cite[Prop.~1.3]{adamek1994locally} and every finite set is the
  finite coproduct of singleton sets (\autoref{suppCoprod}), we can assume wlog
  that $X$ is singleton $X=\{x\}$. For the verification that $X$ is fp, consider
  a supported map $f\colon X\to \colim D$ for a directed diagram $D\colon \D\to
  \Supp{A}$. Since colimit injections are jointly surjective in $\Set$ and thus
  also in $\Supp{A}$, there are $Z\in \D$, $z\in DZ$ with $\inj_Z(z) = f(x)$.
  However, the support of $z$ could possibly contain more than
  $\s{}(f(x))$. Denote this difference by the finite set
  \[
    R := \s{DZ}(z)\setminus  \s{\colim D}(f(x)) \qquad\subseteq A.
  \]
  \begin{itemize}
  \item If $R:=\emptyset$, then $\s{DZ}(z) = \s{\colim D}(f(x))\subseteq \s{X}(x)$
    and we have the desired supported map $g\colon X\to DZ$, $g(x) = z$.
  \item
    By the characterization of colimits in $\Supp{A}$ (\autoref{suppColim}), we know that
    \begin{align*}
      \dispind
      \s{\colim D}(f(x))
      &= \bigcap\set{\s{Y}(y) \mid Y\in \D, \inj_Y(y) = f(x)}
    \end{align*}
    So for every $a\in R$, there must be some $Y_a\in \D$ and $y_a\in DY_a$ with
    \[
      \inj_{Y_a}(y_a) = f(x)
      \text{ and }
      a\not\in \s{DY_a}(y_a).
    \]
    Let $U\in \D$ be the upper bound of $\set{Y_a\mid a\in R}$ in $\D$, which
      exists by assumption. Then $u := D(Y_a\to U)(y_a)$ fulfils
      \[
        \dispind
        \s{U}(u) = \s{\colim D}(f(x)) \subseteq \s{X}(x)
        \text{ and }
        \inj_U(u) = f(x)
      \]
      so we have the desired supported map $g\colon X\to U$, $g(x) = u$.
  \end{itemize}
  The essential uniqueness of the factorization follows on the level of \Set:
  consider two factorizations:
  \[
    \dispind
    g\colon X\to DY
    \quad
    g'\colon X\to DY'
    \quad
    \inj_Y\cdot g = f = \inj_{Y'}\cdot g'
  \]
  Since $g(x)$ and $g'(x)$ are identified in the colimit ($\inj_Y(g(x))$ $=$
  $\inj_{Y'}(g'(x))$) and since $\D$ is directed, by
  \cite[Ex.~1a.2.ii]{adamek1994locally} there is some $Z\in \D$ with
  $Z\ge Y$, $Z\ge Y'$ and \[
    D(Y\to Z)(g(x)) = D(Y'\to Z)(g'(x)).
  \]
  \twnote{}

\item Since $\Supp{A}$ is cocomplete, and since we have characterized the
  finitely presentable objects as the finite supported sets,
  \autoref{itemDirColimSubset} shows that $\Supp{A}$ is lfp.
  \qedhere
\end{enumerate}
\end{proofappendix}

\subsection{Injectivity, Surjectivity, Quasitopoi}
Notions of surjective and injective maps generalize from $\Set$:
\begin{lemma} \label{epimono}
  Monomorphisms in $\Supp{A}$ are precisely the injective supported maps
  and epimorphisms are precisely the surjective supported maps.
\end{lemma}
\begin{proofappendix}{epimono}
  \begin{itemize}
  \item Epi = surjective: Every surjective supported map is an epimorphism
    because the forgetful $\ToSet \colon \Supp{A}\to \Set$ is faithful.
    Conversely, every epimorphism is surjective, because $\ToSet $ is left-adjoint (\autoref{setSubcat}).

  \item Mono = injective: Every injective supported map is a monomorphism 
    because the forgetful $\ToSet \colon \Supp{A}\to \Set$ is faithful. Conversely,
    for a monomorphism $m\colon X\to Y$ in $\Supp{A}$,\twnote{} consider $a,b\in X$ with $m(x) = m(y)$.
    Define \[
      \text{$Z= 1=\{0\}$ with $\s{Z}(0) = \s{X}(a) \cup \s{X}(b)$},
    \]
    hence we have
    supported maps $f_a,f_b\colon Z\to X$ with $f_a(0) =a$, $f_b(0) = b$. Thus,
    $m\cdot f_a = m\cdot f_b$. Since $m$ is monic, we obtain $f_a=f_b$ and so
    $a=b$ as desired.
    \qedhere
  \end{itemize}
\end{proofappendix}

However, not all bijective supported maps are isomorphisms in $\Supp{A}$,
because they may drop atoms. That is, the difference between bijections and isomorphisms
is the following:
\begin{definition}
  A map $f\colon X\to Y$ is called \emph{support"=reflecting} if
  $\s{Y}\cdot f = \s{X}$.
\end{definition}

\begin{lemma} \label{propSuppIso}
  The isomorphisms in $\Supp{A}$ are the support\hyp{}reflecting
  bijective maps.
\end{lemma}
\begin{proofappendix}{propSuppIso}
  If a bijective map $f\colon X\to Y$ is support reflecting,
  then $f^{-1}\colon Y\to X$ fulfils $\s{Y} = \s{X}\cdot f^{-1}$ and hence
  $f^{-1}$ is a supported map; thus $f$ is an isomorphism.

  Conversely, if $f\colon X\to Y$ is an isomorphism, then $f$ is bijective and
  we have a supported map $f^{-1}\colon Y\to X$ with
  \[
    \s{Y}(f(x)) \supseteq \s{X}(f^{-1}(f(x)))
    = \s{X}(x).
  \]
  Thus, $\s{Y}\cdot f= \s{X}$ in total, i.e.~$f$ is support-reflecting.
\end{proofappendix}

\begin{example}
  The unit $\eta_A\colon A\to J \ToSet A$ of the above adjunction $\ToSet \dashv J$ to $\Set$ is a
  bijective supported map but not support-reflecting, because $\s{A}(a) = \set{a}$ and
  $\s{J\ToSet A}(a) = \emptyset$. Thus it is not an
  isomorphism in $\Supp{A}$.
\end{example}

This shows that $\Supp{A}$ is not a topos -- in contrast to $\Set$ and $\Nom$.
As we will see, it is a \emph{quasitopos}~\cite[Def.~2.6.1]{Johnstone02}, which
entails that $\Supp{A}$ is locally cartesian closed and that it has a subobject
classifier for \emph{regular} monomorphisms. The precise definition of
regularity is not relevant here~(but cf.~\cite[Rem 7.76(2)]{joyofcats}), since
it can be nicely characterized via support\hyp{}reflection:

\begin{lemma} \label{classifier}
  A monomorphism is regular iff it is support\hyp{}reflecting.
  Moreover, $t\colon 1\to 2$, $0\mapsto 1$ is a regular\hyp{}subobject classifier,
  i.e.~for every supported set $X$, the support\hyp{}reflecting monomorphisms $m\colon S\to X$ are in
  correspondence to characteristic maps $\chi_S\colon X\to 2$.
\end{lemma}
\begin{proofappendix}{classifier}
  We prove both statements about regular monomorphisms via the following steps:
  \begin{enumerate}
  \item\label{itemSubobjClass} There is a subobject classifier for support-preserving monomorphisms.
  \item Every regular monomorphism is support-preserving.
  \item Every support-preserving monomorphism is regular (using item \ref{itemSubobjClass}).
  \end{enumerate}
  For the verification:

  \begin{enumerate}
  
  \item \label{itemClassifier}
    For a support-preserving monomorphism $m\colon S\to X$,
    We need to define $\chi_S\colon X\to 2$ such that
    \[
      \begin{tikzcd}
        S
        \arrow{r}{m}
        \arrow{d}[swap]{!}
        & X
        \arrow{d}{\chi_S}
        \\
        1
        \arrow{r}{t}
        & 2
      \end{tikzcd}
    \]
    is a pullback square.
  Define $\chi_S\colon X\to 2$ to be the usual characteristic function, i.e.~the
  subobject classifier in $\Set$:
  \[
    \chi_S(x) = \begin{cases}
      1 &\text{if there is some }s\in S\text{ with }m(s) = x\\
      0 &\text{otherwise}.
    \end{cases}
  \]
  Hence, $\chi_S\cdot m = t\cdot !$ since the faithful $\ToSet \colon
  \Supp{A}\to\Set$ sends this to the commutative square of the subobject
  classifier in $\Set$.

  For the verification of the universal property of the pullback, consider a
  cone $(C,c,d)$, i.e.~supported maps $!_C\colon C\to 1$, $d\colon C\to X$ with
  $\chi_S\cdot d = t\cdot \mathord{!_C}$.
  \[
    \begin{tikzcd}
      C
      \arrow[bend right]{ddr}[swap]{!_C}
      \arrow[bend left]{drr}{d}
      \arrow[dashed]{dr}{u}
      \\
      &S
      \arrow{r}{m}
      \arrow{d}[swap]{!}
      & X
      \arrow{d}{\chi_S}
      \\
      & 1
      \arrow{r}{t}
      & 2
    \end{tikzcd}
  \]
  The universal property of the subobject classifier in $\Set$ induces a map
  $u\colon C\to S$ with $m\cdot u = d$ and $!\cdot u = \mathord{!_C}$. All we
  need to verify is that $u$ is a supported map. Since $m$ is
  support-preserving, we have $\s{X}\cdot m = \s{S}$ and thus for all $y\in C$
  \[
    \s{S}(u(y))
    = \s{X}(m(u(y)))
    = \s{X}(d(y))
    \subseteq \s{C}(y)
  \]
  as required.

  The uniqueness of $u$ is clear because $\ToSet \colon \Supp{A}\to\Set$ is faithful.

  \item \label{itemRegularSuppPres}
    Consider a regular monomorphism $m\colon S\to X$, that is, the equalizer of a
    parallel pair of morphisms $f,g\colon X\to Y$:
    \[
      \begin{tikzcd}
        S\arrow{r}{m}
        & X \arrow[shift left=1]{r}{f}
        \arrow[shift right=1]{r}[swap]{g}
        & Y
      \end{tikzcd}
    \]
    By \autoref{suppLim}, the support on the limit $S$ is
    \[
      \s{S}(z)
      \overset{\text{\eqref{supportOnLimit}}}{=}
      \s{X}(m(z)) \cup \s{Y}(f(m(z))) 
      = \s{X}(m(z))
    \]
    where we use that $f$ is a supported map in the second equation.

  \item Given a support-preserving monomorphism $m\colon S\to X$, we have a
    pullback square
    \[
      \begin{tikzcd}
        S
        \arrow{r}{m}
        \arrow{d}[swap]{!}
        & X
        \arrow{dl}[swap]{!}
        \arrow{d}{\chi_S}
        \\
        1
        \arrow{r}{t}
        & 2
      \end{tikzcd}
    \]
    Since $1$ is the terminal object, we have that $m$ is the equalizer of
    $\chi_S$ and $t\cdot !$:
    \[
      \begin{tikzcd}
        S\arrow{r}{m}
        & X \arrow[shift left=1]{r}{\chi_S}
        \arrow[shift right=1]{r}[swap]{t\cdot !}
        & 2
      \end{tikzcd}
    \]
    (Note that this is just the argument that if a monomorphism $m$ has a subobject
    classifier, then it is a regular monomorphism).
    \qed\qedhere
  \end{enumerate}
\end{proofappendix}

In summary, for every set $A$, $\Supp{A}$ is (co)complete, cartesian
closed, and locally finite, and we can express many of the above mentioned
properties in one line:

\begin{theorem}
  \label{suppQuasiTopos}
  $\Supp{A}$ is a quasitopos (for both
  \cite[Def.~2.6.1]{Johnstone02} and \cite[Def.~28.7]{joyofcats}).
\end{theorem}
Thus, $\Supp{A}$ constitutes a rich basis to study algebraic theories on it, such as
nominal sets.
\begin{proofappendix}{suppQuasiTopos}
  There exist slightly different definitions of \emph{quasitopos} in the
  literature, which vary in the class of monomorphisms considered.

  There is an entire hierarchy of monomorphisms classes~\cite[Rem
  7.76(2)]{joyofcats}, containing extremal, strong, and regular monomorphisms,
  among others. Details are not important for the present work, because the
  entire hierarchy collapses into a single notion whenever regular
  monomorphisms are closed under composition~\cite[Prop.~14.14]{joyofcats}.
  This is the case in $\Supp{A}$, since regular monos are the
  support\hyp{}reflecting monos (\autoref{classifier}), a notion clearly closed
  under composition. Hence, we only need to distinguish monomorphisms and
  support\hyp{}reflecting monomorphisms in $\Supp{A}$.

  For the verification that $\Supp{A}$ is a quasitopos, we consider
  the textbooks definitions by Adámek~\etal{} and Johnstone:

  \subparagraph{Adámek, Herrlich, and Strecker}~\cite[Def.~28.7]{joyofcats} require that a quasitopos
  \begin{itemize}
  \item is finitely cocomplete,
  \item is cartesian closed, and
  \item has representable extremal partial morphisms.
  \end{itemize}
  We do not recall the last condition~\cite[Def.~28.1]{joyofcats} in full
  generality here. The condition simplifies in $\Supp{A}$ because extremal
  monomorphisms are the support\hyp{}reflecting ones and because $\Supp{A}$ has
  all pullbacks. Instantiated to our setting, it suffices to verify that
  for all support\hyp{}reflecting monomorphisms $m\colon X\to Y$ and supported
  maps $f\colon X\to B$, there exists a unique map $g\colon Y\to B+1$ such that
  \[
    \begin{tikzcd}
      X
      \arrow[>->]{r}{m}
      \arrow{d}[swap]{f}
      & Y
      \arrow{d}{g}
      \\
      B
      \arrow{r}{\inl}
      & B+1
    \end{tikzcd}
  \]
  is a pullback square.
  We can simply define
  \[
    g\colon Y\to B+1
    \qquad
    g(y) = \begin{cases}
      \inl(f(x)) &\text{if there exists }x\in X\text{ with }m(x) = y\\
      \inr(0) &\text{otherwise}.
    \end{cases}
  \]
  Since $m$ is injective, $g$ is well-defined and since $m$ is
  support\hyp{}reflecting, $g$ is a supported map. Pullbacks are finite limits,
  so formed as in \Set. Hence, the uniqueness of $g$ and the verification of
  the pullback's universal property is just as in \Set~\cite[Ex.~28.2(1)]{joyofcats}.

  \subparagraph{Johnstone}~\cite[Def.~2.6.1]{Johnstone02}
  requires a quasitopos to be
  \begin{itemize}
    \item finitely cocomplete (called \textqt{cocartesian} in \cite{Johnstone02}),
    \item locally cartesian closed (a property slightly stronger than cartesian closed),
    \item equipped with a subobject classifier for strong monomorphisms (called \textqt{cocovers} in \cite{Johnstone02}).
  \end{itemize}

  For the verification, we have already seen that $\Supp{A}$ is finitely
  cocomplete and that $\Supp{A}$ has a regular-subobject classifier $2$
  (\autoref{classifier}), and that strong and regular monomorphisms
  coincide in $\Supp{A}$.

  The remaining property to verify is locally cartesian closedness, which means
  that every slice category $\Supp{A}/I$ (for arbitrary $I\in \Supp{A}$) is cartesian closed.
  Recall that an object of $\Supp{A}/I$ is a
  pair $(B,b)$ where $B\in \Supp{A}/I$ and $b\colon B\to I$. The morphisms in
  $\Supp{A}/I$ are commutative triangles ($f\colon (B,b)\to (C,c)$ is $f\colon
  B\to C$ with $c\cdot f = b$). For example, the slice category $\Supp{A}/1 =
  \Supp{A}$ is indeed cartesian closed (\autoref{suppCCC}). For a general supported set $I$,
  an object $(B,b)\in \Supp{A}/I$ consists of an $I$-indexed supported set $B$.
  By writing $B_i := \{x\in B\mid b(x) = i\}$ for component $i\in I$, we have
  that $(B,b)$ is equivalently a family of supported sets
  \[
    (B_i)_{i\in I}\text{ with }\s{B_i}(x) \supseteq \s{I}(i)
    \text{ for all }x\in B_i.
  \]
  Hence, $\Supp{A}/I$ is equivalent to a product of slice categories:
  \[
    \Supp{A}/I \cong \prod_{i\in I} \Supp{A}/\set{i}
  \]
  In a slice category $\Supp{A}/\set{i}$ for a singleton $\set{i}$ (for $i\in
  I$), all elements of all supported sets have at least support $\s{I}(i)$, and
  the morphisms $f\colon B\to C$ in $\Supp{A}/\set{i}$ are just ordinary
  supported maps $f\colon B\to C$ in $\Supp{A}$ since $\set{i}$ is singleton. Thus, $\Supp{A}/\set{i}\cong \Supp{A\setminus
    \s{I}(i)}$, which is cartesian closed by \autoref{suppCCC}.
  Thus, the category
  \[
    \Supp{A}/I \cong \prod_{i\in I} \Supp{A}/\set{i}
    \cong \prod_{i\in I} \Supp{A\setminus\set{i}}.
  \]
  is the (possibly infinite) product of cartesian
  closed categories and therefore cartesian closed itself.
\end{proofappendix}

\section{Monadicity of Nominal $M$-Sets}
\label{secMonadicity}
If a category is \emph{monadic}, then intuitively, it is a well-behaved class of
algebras. If so, it becomes applicable for many results and constructions,
e.g.~regarding representation and the generalized powerset construction.
Algebraic theories have been studied extensively throughout the decades and are
in correspondence to monads: given a monad $T$ on a category $\C$, e.g.~\Set, its
\emph{Eilenberg-Moore category} $\EM(T)$ contains the models of the algebraic
theory defined by $T$ (see e.g.~\cite[10.3]{awodey2010category}).
The forgetful functor $U\colon \EM(T)\to \C$ is a right-adjoint functor and its
left-adjoint $F\colon \C\to \EM(T)$ sends an object $X\in \C$ (e.g.~a set of
generators) to
the \emph{free algebra} on $X$:
\begin{equation}
  F\colon \Supp{A}\to \EM(T)
  \quad
  \dashv
  \quad
  U\colon \EM(T)\to\Supp{A}
  \qquad
  \text{(for $\C=\Supp{A}$)}
  \label{adjointEM}
\end{equation}
The category of $M$-sets is monadic over \Set, but $\Nom(M)$ fails
to be monadic over \Set in the instances of interest (\autoref{MonList}).
The reason for this is that infinite products in $\Nom(M)$ are different than in
\Set~\cite{pittsbook}, so the forgetful functor $\Nom(M)\to \Set$ is not right-adjoint and therefore not monadic.
As we will show, $\Nom(M)$ is still monadic, but \emph{over supported
  sets}.

\begin{definition}
  A functor $U\colon \D \to \C$ is called \emph{monadic (over $\C$)} if there is
  a monad $T$ on $\C$ such that $\D$ is $\EM(T)$ and $U$ is the forgetful
  functor.
\end{definition}

We first show that $U\colon \Nom(M)\to \Supp{A}$ is right-adjoint
and then that it is monadic. Then, we can view nominal sets as algebras over $\Supp{A}$. There,
the algebraic operations come from the following (supported) set:
\begin{definition} \label{restmon}
  For $M\le A^A$ and a set $S\subseteq A$, put
  $[m]_{S}$ for the $\resteq{S}$-equivalence class of $m$
  (cf.~\autoref{defRestEq})
  and
  \(
    \restmon[M]{S} = \set{[m]_{S}\mid m\in M}
  \) for the supported set of equivalence classes
  with 
  $\s{\restmon[M]{S}}([m]_S) := m[S] = \set{m(a)\mid a\in S}$.
\end{definition}
\begin{proofappendix}[Details for]{restmon}
  Note that $\restmon[M]{S}$ is the image of the composition
  $M\monoto A^A\epito A^S$ of the submonoid inclusion $M\monoto A^A$ and the
  restriction of maps $A^A\epito A^S$ to $S\subseteq A$:
  \begin{equation}
    \begin{tikzcd}
      M
      \arrow[>->]{r}
      \arrow[->>]{d}
      & A^A
      \arrow[->>]{d}
      \\
      \restmon[M]{S}
      \arrow[>->]{r}
      & A^S
    \end{tikzcd}
    \quad\text{(in sets)}
    \label{eqImgMS}
  \end{equation}
\end{proofappendix}
We can consider $\restmon{S}$
either as equivalence classes of $M$-elements that are identical on $S$
or alternatively as special maps $S\to A$ that are obtained by restricting some $m\in
M\subseteq A^A$ to
$S\subseteq A$.
Intuitively, $\restmon{S}$ is the free nominal set on one generator with support
$S$. Hence, the free nominal $M$-set over a supported set $X$ is the union of
multiple $\restmon{S}$:

\begin{definition} \label{freeNominal} 
  Fix the functor
  $\Monad\colon \SSet{A}\to\SSet{A}$
  by
  \(
  \Monad X = \coprod_{x\in X} \restmon[M]{\s{}(x)}.
  \)
\end{definition}
\begin{proofappendix}[Verification of]{freeNominal}
  The functor $\Monad\colon \SSet{A}\to\SSet{A}$ sends $X$
  to
  \[
  \Monad X =
  \coprod_{x\in X} \restmon[M]{\s{}(x)}
  = \{([m]_{\s{X}(x)},x)\mid x\in X\}
  \]
  and a supported map $f\colon X\to Y$ to the supported map
  \begin{align*}
    &\Monad f\colon  \Monad X\to \Monad Y
    &
    \\
    &(\Monad f)([m]_{\s{X}(x)},x) = ([m]_{\s{Y}(f(x))}, f(x)).
  \end{align*}

  For the verification that $\Monad$ is a functor, consider a supported map
  $f\colon X\to Y$ and $([m]_{\s{}(x)},x) \in \Monad X$. The map $\Monad f$ is
  well-defined, because
  \[
    \s{Y}(f(x))\subseteq \s{X}(x)
    ~~\text{ implies }~~
    [m]_{\s{X}(x)}
    \subseteq [m]_{\s{Y}(f(x))},
  \]
  so if $[m]_{\s{}(x)} = [m']_{\s{}(x)}$, then
  also $[m]_{\s{}(f(x))} = [m']_{\s{}(f(x))}$. It is straightforward to see that
  $\Monad(-)$ preserves identities and composition.
\end{proofappendix}
$\Monad$ is the monad that will define nominal $M$-sets as an algebraic theory.
Written with elements, we have $\Monad X = \{([m]_{\s{X}(x)},x)\mid x\in
X\}$ with \( \s{\Monad X}([m]_{\s{}(x)},x) =
m[\s{}(x)] \). For a supported set $X$, every such element $([m]_{\s{}(x)},x)$ of the nominal set
$\Monad X$ is equivalently an element $x\in X$ together with an
$|\s{}(x)|$-tuple of atoms. Of course, such an equivalence class $[m]_{\s{}(x)}$
is not an arbitrary tuple of elements of $A$ but only one that is obtained from
restricting some $m\in M\subseteq (A\to A)$ to $\s{X}(x)\subseteq A$. For
example, for the equality symmetry $M := \Perm(\A)$, such a $t\in
\restmon[M]{\s{X}(x)}$ is a tuple of \emph{distinct} elements of $A$.

\begin{proposition}
  \label{freeToNom}
  $\Monad X$ with the $M$-set action
  \(
    \ell\cdot ([m]_{\s{}(x)},x)
    = ([\ell\cdot m]_{\s{}(x)},x)
  \)
  is nominal and gives rise to a functor
  \(
    F\colon \SSet{A}\to
    \Nom(M)
  \),
  \(
    F X = \Monad X.
  \)
\end{proposition}
\begin{proofappendix}{freeToNom}
  We first show a couple of auxiliary results:
  \begin{lemma}
    \label{restEqCongr}
    For every $S\subseteq A$,
    $\resteq{S}$ is an equivalence relation and a congruence w.r.t.~multiplication
    from the left. That is,
    \(
    m\resteq{S}m'\text{ implies }\ell\cdot m \resteq{S}\ell\cdot m'
    ~\text{for all }m,m',\ell\in M.
    \)
  \end{lemma}
  \begin{proof}
    By definition, $\resteq{S}$ is reflexive, symmetric, and transitive.

    We now verify that $m\resteq{S}m'$ implies $\ell\cdot m \resteq{S}\ell\cdot m'$ for all $\ell,m,m'\in M$.
    If $m\resteq{S} m'$, then $m(a) = m'(a)$ for all $a\in S$. Hence, $\ell(m(a))
    = \ell(m'(a))$ for all $a\in S$, i.e.~$\ell\cdot m \resteq{S} \ell\cdot m'$.
  \end{proof}

  \begin{lemma}
    \label{restmonNominal}
    $\restmon[M]{S}$ is a nominal $M$-set via
    $\ell\cdot [m]_S := [\ell\cdot m]_S$ and $m[S]$ supports $[m]_S$.
  \end{lemma}
  \begin{proof}
    Since $\resteq{S}$ is a congruence for left-multiplication (\autoref{restEqCongr}),
    the $M$-set structure on
    $\restmon[M]{S}$ by $\ell\cdot [m]_S := [\ell\cdot m]_S$ is well-defined. For the verification that $m[S]$ supports $[m]_S$, take
    $\ell,\ell'\in M$ with $\ell \resteq{m[S]} \ell'$. For all $a\in S$ we have
    $m(a) \in m[S]$ and so $\ell(m(a)) = \ell'(m(a))$. Thus,
    $[\ell\cdot m]_S = [\ell'\cdot m]_S$ and
    \[
      \ell\cdot [m]_S
      = [\ell\cdot m]_S
      = [\ell'\cdot m]_S
      = \ell\cdot [m]_S.
      \tag*{\qedhere}
    \]
  \end{proof}

  By \autoref{restmonNominal}, $\restmon[M]{\s{}(x)}$ is a nominal $M$-set, and
  hence the disjoint union
  \[
    \Monad X = \coprod_{x\in X}\restmon[M]{\s{}(x)}
  \]
  is also a nominal $M$-set.

  For supported map $f\colon X\to Y$, $\Monad f\colon \Monad X\to \Monad Y$ is
  equivariant because
  \begin{align*}
    \ell\cdot (\Monad f)([m]_{\s{X}(x)},x)
    &= \ell\cdot ([m]_{\s{Y}(f(x))},f(x))
    \\
    &= ([\ell\cdot m]_{\s{Y}(f(x))},f(x))
      \\ &
      = (\Monad f)([\ell\cdot m]_{\s{X}(x)},x)
           \tag*{\qedhere}
  \end{align*}
\end{proofappendix}

Note that there is no statement about the least finite support of
$([m]_{\s{}(x)},x)\in \Monad X$. In the proof, we show that it is supported by
$m[\s{}(x)]$, but the least support might be smaller. Nevertheless, the
existence of finite supports in $\Monad X$ suffices to show the adjunction $F\dashv U$:

\begin{proposition} \label{thmFree}
  If $M\le A^A$ admits least supports, then
  $F\colon \Supp{A}\to\Nom(M)$ is left-adjoint to
  $U\colon \Nom(M)\to \Supp{A}$
  with unit $\eta_X\colon X\to UF X$, $\eta_X(x) = ([\id_A]_{\s{}(x)}, x)$.
\end{proposition}
\begin{proofappendix}{thmFree}
  The proposed unit $\eta_X$ is a supported map $\eta_X\colon X\to UFX$
  because
  \begin{align*}
    \s{UFX}(\eta_X(x))
    &= \s{UF X}([\id_A]_{\s{}(x)}, x)
      \\ &
    = \supp{}([\id_A]_{\s{}(x)}, x)
      \tag{\autoref{forgetful}}
    \\
    &\subseteq
    \id_A[\s{X}(x)] 
    \tag{\autoref{restmonNominal}}
    = \s{X}(x).
  \end{align*}
  For the universal property, consider a supported map $f\colon X\to UY$ (in
  $\SSet{A}$) for some nominal $M$-set $Y$. We need to show that there is a
  unique $g\colon \Monad X\to Y$ in $\Nom(M)$ with $g([\id_A]_{\s{}(x)},x) =
  g(\eta_X(x)) = f(x)$:
    \[
      \begin{tikzcd}
        U(\Monad X)
        \arrow{r}{Ug}
        & UY
        \\
        X
        \arrow{u}{\eta_X}
        \arrow{ur}[swap]{f}
      \end{tikzcd}
    \]
    Define $g$ by
    \[
      g([m]_{\s{}(x)},x) := m\cdot f(x)
    \]
    which clearly fulfils $g(\eta_X(x)) = f(x)$. For well-definedness,
    consider $m \resteq{\s{}(x)} m'$, and thus $m(a) = m'(a)$ for all
    $a\in \s{X}(x)$. Since $M$ admits least supports, we have
    \[
      \supp{Y}(f(x)) =
      \s{UY}(f(x)) \subseteq \s{X}(x),
    \]
    and so
    $m(a) = m'(a)$ for all $a\in \supp{Y}(f(x))$. Thus, 
    $m\cdot f(x) = m'\cdot f(x)$ (\autoref{defNominal}). Clearly, $g$ is
    equivariant:
    \begin{align*}
      g(\ell\cdot ([m]_{\s{}(x)}, x))
      &= g([\ell\cdot m]_{\s{}(x)}, x)
      = \ell\cdot m\cdot f(x)
        \\ &
      = \ell\cdot g([m]_{\s{}(x)}, x)
      \qquad\text{ for all }\ell\in M.
    \end{align*}
    For uniqueness, consider another equivariant $g'\colon \Monad X\to Y$ with
    $g'([\id_A]_{\s{}(x)}, x) = g'(\eta_X(x)) =  f(x)$. By equivariance, we
    obtain
    \begin{align*}
      g'([m]_{\s{}(x)}, x)
      &= g'([m\cdot \id_A]_{\s{}(x)}, x)
      = g'(m\cdot ([\id_A]_{\s{}(x)}, x))
      \\ &
      = m\cdot g'([\id_A]_{\s{}(x)}, x)
      \\ &
      = m\cdot f(x)
      = g([m]_{\s{}(x)}, x),
    \end{align*}
    i.e.~$g' = g$.
\end{proofappendix}

This adjunction shows that every nominal $M$-set is an Eilenberg-Moore algebra
and that every supported set generates a free nominal $M$-set satisfying a
universal mapping property. The remaining direction to prove is that every
Eilenberg-Moore algebra for $\Monad$ is indeed a nominal set, or concretely,
that the adjunction is monadic. For doing so, we impose a property on the
monoid $M$ of interest, which will not only be used in the monadicity proof but
also help us to characterize the least finite supports of the free nominal sets:

\begin{notheorembrackets}
\begin{definition}[{\cite[Def.~9.6]{BojanczykEA14}}]\label{defLockFree}
  A monoid $M\le A^A$ is called \emph{fungible} if for all finite $R\subseteq A$ and
  $a\in A\setminus R$, there is some
  $\ell\in M$ with $\ell \resteq{R} \id_A$ and $\ell(a) \neq a$.
\end{definition}
\end{notheorembrackets}
\begin{proofappendix}[Equivalent formulation of]{defLockFree}
  In proofs, we will use the contrapositive of \autoref{defLockFree}:
\begin{lemma} \label{defLockFreeContra}
  $M\le A^A$ is fungible iff for all $a\in A$ and
  finite $R\subseteq A$ we have that
  \[
    (\forall \ell\in M, \ell\resteq{R}\id_A\colon \ell(a) = a)
    \quad\text{ implies }\quad
    a\in R.
  \]
\end{lemma}
\begin{proof}
  To see that this is indeed equivalent, we have the following chain of
  equivalences for all $a\in A$ and finite $R\subseteq A$:
  \begin{itemize}
  \item[]
    $a\not\in R$
    implies
    there exists
    $\ell\in M$ with $\ell \resteq{R} \id_A$ and $\ell(a) \neq a$.

  \item[$\Leftrightarrow$]
    $a\not\in R
    \quad \Longrightarrow\quad
    \exists \ell\in M: \ell\resteq{R} \id_A \wedge \ell(a) \neq a$

  \item[$\Leftrightarrow$]
    $
    (\neg\exists \ell\in M: \ell\resteq{R} \id_A \wedge \ell(a) \neq a)
    \quad \Longrightarrow\quad
    a\in R
    $

  \item[$\Leftrightarrow$]
    $
    (\forall \ell\in M: \neg(\ell\resteq{R} \id_A) \vee \ell(a) = a)
    \quad \Longrightarrow\quad
    a\in R
    $

  \item[$\Leftrightarrow$]
    $
    (\forall \ell\in M, \ell\resteq{R} \id_A\colon \ell(a) = a)
    \quad \Longrightarrow\quad
    a\in R
    $
    \qedhere
  \end{itemize}
\end{proof}
\end{proofappendix}

Intuitively, the condition expresses that the atoms in the support can be
renamed independently from each other: if an element is supported by $R\cup\{a\}$, we can
always rename $a$ to something fresh while keeping the rest of the support fixed.

\begin{example}
  \label{MonListLockFree}
  All the leading examples~(Ex.~\ref{MonList}) have fungible
  monoids.
  \begin{itemize}[topsep=0pt]
  \item For $M:=\Perm(\A)$ and $M:=\Fin(\A)$, consider a finite $R\subseteq \A$
    and $a\fresh R$. Then for $b\fresh a, R$, the permutation $\ell = (a\,b)$ fulfils
    the desired property $(a\, b) \resteq{R}\id_\A$.

  \item For $M:=\Aut(\Q,<)$, the verification uses a notion called
  homogeneity~\cite[Lemma 5.2]{VenhoekEA19}.
  \end{itemize}
\end{example}
\begin{proofappendix}[Verification of]{MonListLockFree}
  For plain $\Nom$ and renaming nominal sets, one simply picks some $b\in A$
  fresh for $R$ and $\{a\}$ and puts $\ell = (a\, b)$, i.e.~$\ell(a) = b$,
  $\ell(b) = a$, and $\ell(x) =x $ for $x\not\in \{a,b\}$.

  For the order symmetry, a more involved argument is necessary.
  Venhoek~\etal~\cite[Lemma 5.2]{VenhoekEA19} show a property called
  \emph{homogeneity}: for any two finite
  $C\subseteq \Q$, $C\subseteq \Q'$, if $|C| = |C'|$ then there is a $\pi \in
  \Aut(\Q,<)$ with $\pi[C] = C'$.

  Apply this for
  \[
    b := a + \frac{1}{2} \cdot \min_{x \in R} |a - x|
    \qquad
    C := R \cup \{a\}
    \qquad
    C' := R \cup \{b\}
  \]
  Since $a\not\in R$, we have $b\neq a$ and also $b\not\in R$. Hence, $|C| =
  |C'| = |R|+1$ and homogeneity provides us with $\pi \in \Aut(\Q,<)$ with
  $\pi[C] = C'$. Since $\pi$ is monotone and $b$ is closer to $a$ than any
  element in $R$, we necessarily have $\pi(x) = x$ for all $x\in R$ and $\pi(a)
  = b$.

  Note that for general monoids $M$ (i.e.~$A$ instead of $\Q$ and $M$ instead of
  ${\Aut(\Q,<)}$), homogeneity is a different notion to fungibility. Take for
  instance $A := \Sigma\times \A$ for a finite set $\Sigma$, and let $M$ be the
  finite bijections on $A$ that fix $\Sigma$:
  \[
    M := \{ f\in \Perm(\Sigma\times \A)\mid
    \pi_1(f(\sigma,a)) = \sigma
    \text{ for all }(\sigma,a)\in A
    \}
  \]
  Then, $M$ is clearly fungible: given $(\sigma,a)\in A\setminus R$, just
  consider the transposition
  $\ell = ((\sigma,a)\ (\sigma,b))$ for some fresh $b$. However, $M$ does not
  fulfil homogeneity. Simply put $C = \{(\sigma,a)\}$, $C'=\{(\sigma,b)\}$ for
  distinct $a,b\in \A$ and some $\sigma \in \Sigma$; then, there is no $\pi\in
  M$ with $\pi[C] = C'$.
\end{proofappendix}

\begin{lemma}
  \label{restmonSupp}
  If $M$ is fungible, then $\Monad X$ is a nominal $M$-set with
  $\supp{}([m]_{\s{}(x)},x) = m[\s{}(x)]$ for every $m\in M$ and $x\in X$.
\end{lemma}
\begin{proofappendix}{restmonSupp}
  After \autoref{restmonNominal} it remains to show
  that $m[S] \subseteq R$ for every finite $R\subseteq A$ that supports $[m]_S$.

  Consider an arbitrary $\ell\in M$ with $\ell\resteq{R} \id_A$. Since $R$
  supports $[m]_S$ (\autoref{defNominal}), this implies $\ell\cdot [m]_S =
  \id_A\cdot [m]_S$. Thus, $[\ell\cdot m]_S = [m]_S$ and so $\ell(m(a)) = m(a)$
  for all $a\in S$. So $M$ and $R$ fulfil $\forall \ell\in M,
  \ell\resteq{R}\id_A\colon \ell(m(a)) = m(a)$ for all $a\in S$. Hence by
  \autoref{defLockFreeContra}, we obtain $m(a) \in R$ for all $a\in S$. In other
  words, $m[S] \subseteq R$.
\end{proofappendix}
\noindent
We prove the adjunction to be monadic via
Beck's theorem~\cite[Sec.~VI.7]{lane1998categories},
which will provide us with the monadicity of the leading examples of nominal sets
by instantiating $M$.
\begin{theorem} \label{NomEMBeck}
  $U\colon \Nom(M)\to\SSet{A}$ is monadic and $\Nom(M) = \EM(\Monad(-))$,
  for every fungible $M$ admitting least supports.
\end{theorem}
\begin{proofappendix}{NomEMBeck}
  We have seen in \autoref{thmFree} that $U\colon \Nom(M)\to \SSet{A}$ is
  right-adjoint if $M$ admits least supports. For the monadicity, we use Beck's theorem
  (see e.g.~\cite[Section VI.7]{lane1998categories}):
  \begin{theorem}[Beck's theorem]
    For every right-adjoint functor $U\colon \D\to \C$ the following are equivalent:
    \begin{enumerate}[topsep=0pt]
    \item $U\colon \D\to \C$ is monadic.
    \item The functor $U\colon \D\to C$ creates split coequalizers: concretely,
      if $f,g\colon X\to Y$ in $\D$ 
      and morphisms $e,r,t$ in $\C$ make
      \[
        \begin{tikzcd}
          UY
          \arrow{r}{t}
          \arrow[shiftarr={yshift=6mm}]{rr}{\id_{UY}}
          \arrow{d}[swap]{e}
          & UX
          \arrow{r}{Uf}
          \arrow{d}{Ug}
          & UY
          \arrow{d}{e}
          \\
          E
          \arrow[shiftarr={yshift=-6mm}]{rr}{\id_{E}}
          \arrow{r}{r}
          & UY
          \arrow{r}{e}
          & E
        \end{tikzcd}
        \quad\text{in }\C
      \]
      commute, then $f,g$ have a coequalizer $e'\colon Y\to E'$ in $\D$ with $Ue'
      = e$.
    \end{enumerate}
  \end{theorem}

  Also, we will use the following result on the $\supp{}$ maps:
  \begin{notheorembrackets}
  \begin{lemma}[{\cite[Lem.~11]{GabbayH08}}] \label{suppRenameIncl}
    If $(X,\cdot)$ has least supports, then we have
    for all $m\in M$ and $x\in X$:
    \begin{enumerate}[topsep=0pt]
    \item $\supp{X}(m\cdot x) \subseteq m\cdot \supp{X}(x)$.
    \item If $m$ is invertible, then
      $\supp{X}(m\cdot x) = m\cdot \supp{X}(x)$.
    \end{enumerate}
  \end{lemma}
  \end{notheorembrackets}
  \begingroup
  \begin{proof}[Proof of \autoref{suppRenameIncl}]
    The proof for $M:= \Fin(A)$ can be found in \cite[Lem.~11]{GabbayH08} and
    straightforwardly adapts to arbitrary monoids $M$:
    \begin{enumerate}
    \item We show that $m[\supp{X}(x)]$ is a support of $m\cdot x$. Hence, consider
      $k,k'\in M$ with $k\resteq{m[\supp{X}(x)]} k'$.
      Thus, $k\cdot m \resteq{\supp{X}(x)} k'\cdot m$ and so
      $(k\cdot m) \cdot x=(k'\cdot m)\cdot x$ (since $\supp{X}(x)$ supports $x$).
      So $k\cdot (m\cdot x) = k'\cdot (m\cdot x)$, showing that $m[\supp{X}(x)]$
      supports $m\cdot x$.

    \item Let $m^{-1}\in M$ be the (left and right) inverse of $m$. Then we
      verify the other inclusion direction:
      \begin{align*}
        m\cdot \supp{X}(x)
        &= m\cdot \supp{X}(m^{-1}\cdot m\cdot x)
          \subseteq m\cdot m^{-1}\cdot \supp{X}(m\cdot x)
            \tag{by item 1}
        \\ &
            = \supp{X}(m\cdot x)
             \tag*{\qedhere}
      \end{align*}
    \end{enumerate}
  \end{proof}
  \endgroup

  \subparagraph*{Main proof of \autoref{NomEMBeck}.}
  Let us verify that $U\colon \Nom(M)\to\Supp{A}$ reflects split coequalizers:
  Consider a parallel pair of equivariant maps $f,g\colon X\to Y$ in $\Nom(M)$
  and consider supported maps $e,r,t$ such that
  \begin{equation}
    \label{eqSplitCoeqSupp}
    \begin{tikzcd}
      UY
      \arrow{r}{t}
      \arrow[shiftarr={yshift=6mm}]{rr}{\id_{UY}}
      \arrow{d}[swap]{e}
      & UX
      \arrow{r}{Uf}
      \arrow{d}{Ug}
      & UY
      \arrow{d}{e}
      \\
      E
      \arrow[shiftarr={yshift=-6mm}]{rr}{\id_{E}}
      \arrow{r}{r}
      & UY
      \arrow{r}{e}
      & E
    \end{tikzcd}
    \quad\text{in }\SSet{A}
  \end{equation}
  commutes.

  \proofstep{Monoid action $(E,\cdot)$}
  First, define an $M$-set structure on $E$ using the nominal structure on $Y$:
  \[
    m\cdot x := e(\underbrace{m\cdot r(x)}_{\text{in }Y})
    \qquad
    \text{for all }m\in M, x\in E
  \]
  This definition implies that
  \begin{equation}
    m\cdot e(y) = e(m\cdot y)
    \qquad
    \text{for all }m\in M, y\in Y,
    \label{eqEdef}
  \end{equation}
  because we have
  \begin{align*}
    m\cdot e(y)
    &= e(m\cdot r(e(y)))
      \tag{Def.}
    \\
    &= e(m\cdot Ug(t(y)))
      \tag{$r\cdot e=Ug\cdot t$~\eqref{eqSplitCoeqSupp}}
    \\
    &= e(Ug(\underbrace{m\cdot t(y)}_{\text{in }X}))
      \tag{$g$ equivariant}
    \\
    &= e(Uf(m\cdot t(y)))
      \tag{$e\cdot Ug=e\cdot Uf$~\eqref{eqSplitCoeqSupp}}
    \\
    &= e(m\cdot Uf(t(y)))
      \tag{$f$ equivariant}
    \\
    &= e(m\cdot y).
      \tag{$Uf\cdot t = \id_{UY}$~\eqref{eqSplitCoeqSupp}}
  \end{align*}
  Since $e$ is surjective, the defined $M$-set action 
  on $E$ fulfils the required axioms:
  \begin{align*}
    \id_A\cdot e(y)
    &= e(\id_A \cdot y)
    = e(y)
    \\
    (k\cdot m)\cdot e(y)
    &= e((k\cdot m)\cdot y)
    = e(k\cdot (m\cdot y))
      \\ &
      = k\cdot e(m\cdot y)
      = k\cdot (m\cdot e(y)).
  \end{align*}
  Also, equation \eqref{eqEdef} implies that $e$ is an equivariant map. By
  \autoref{equivSupport}, every element of $E$ is finitely supported. Hence,
  $(E,\cdot)$ is a nominal $M$-set, and $e\colon (Y,\cdot)\to (E,\cdot)$ is an
  equivariant map (i.e.~it is in $\Nom(M)$) with $e\cdot f = e\cdot g$.
  For clarity, we explicitly write $(E,\s{E})$ for the
  originally given supported set and $(E,\cdot)$ to denote the constructed nominal set.

  \proofstep{Verification of $U(E,\cdot) = E$}
  We need to verify that $U(E,\cdot) = E$ and in particular $\s{U(E,\cdot)} = \s{E}$. Since $M$,
  admits least supports, we have $\s{U(E,\cdot)}= \supp{(E,\cdot)}$
  (\autoref{forgetful}). We show the two inclusions of $\s{E}(x) =
  \supp{(E,\cdot)}(x)$ for all $x\in E$ separately:
  \begin{itemize}
  \item For the proof of $\s{E}(x) \supseteq \supp{(E,\cdot)}(x)$, we verify
    \begin{align*}
      \supp{(E,\cdot)}(x)
      &=
        \supp{(E,\cdot)}(e(r(x)))
      \\
      &\subseteq 
        \supp{Y}(r(x))
        \tag{$e$ equivariant, \autoref{equivSupport}}
      \\
      &=
        \s{UY}(r(x))
        \tag{Def.~$U$, \autoref{forgetful}}
      \\
      &  \subseteq \s{E}(x)
        \tag{$r$ supported map}
    \end{align*}
  \item For the proof of $\s{E}(x) \subseteq \supp{(E,\cdot)}(x)$, consider $a\in \s{E}(x)$.
    We use the contrapositive formulation of $M$ being fungible (\autoref{defLockFreeContra})
    for
    \[
      R:=\supp{(E,\cdot)}(x) \cup (\s{E}(x)\setminus\{a\}).
    \]
    Hence, we first show that the condition of \autoref{defLockFreeContra}
    applies, namely that
    \[
      \forall m\in M\colon \text{if }
      m\resteq{R} \id_A
      \text{ then }m(a) = a.
      \tag{\(*\)}
    \]
    Consider $m\in M$ with $m\resteq{R} \id_A$.
    By definition, $R\supseteq \supp{(E,\cdot)}(x)$ supports $x$,
    hence~$x=m\cdot x$ and
    \begin{align*}
      a\in \s{E}(x)
      &= \s{E}(m\cdot x) = \s{E}(e(m\cdot r(x)))
        \tag{Def.~$m\cdot x$}
      \\
      &\subseteq \s{UY}(m\cdot r(x))
        \tag{$e$ supported map}
      \\ &
        = \supp{Y}(m\cdot r(x))
        \tag{$\s{UY}=\supp{Y}$ by \autoref{forgetful}}
      \\ &
        \subseteq m[\supp{Y}(r(x))]
        \tag{\autoref{suppRenameIncl}}
      \\ &
        = m[\s{UY}(r(x))]
        \subseteq m[\s{E}(x)]
           \tag{$r$ supported map}
    \end{align*}
    Hence, $a\in m[\s{E}(x)]$, so there is some $b\in \s{E}(x)$ with $m(b) = a$.
    \begin{itemize}
    \item If $b\in R$, then $m(b) = \id_A(b)$ since $m\resteq{R}\id_A$, and so
      $a=b$.
    \item If $b\not\in R$, then in particular $b\not\in \s{E}(x)\setminus\{a\}$.
      Together with $b\in \s{E}(x)$, this implies $b=a$.
    \end{itemize}
    In any case, we have deducted $b=a$ and $m(a) = m(b) = a$ the conclusion of $(*)$.
    Having proven $(*)$, fungibility of $M$ provides us with
    \[
      a\in R = \supp{(E,\cdot)}(x) \cup (\s{E}(x)\setminus\{a\}).
    \]
    Since $a$ is clearly not in the right-hand disjunct, it must be in the left
    disjunct $a\in \supp{(E,\cdot)}(x)$.
  \end{itemize}
  Both inclusions together show that $\s{E}=\supp{(E,\cdot)}$, and so
  $U(E,\cdot) = E$ and automatically $Ue = e$.

  \proofstep{Universal Property of $(E,\cdot)$} It remains to show that $e\colon
  Y\to (E,\cdot)$ is indeed the coequalizer of $f$ and $g$ in $\Nom(M)$. Since
  $e\cdot Uf = e\cdot Ug$ in $\SSet{A}$, we also have $e\cdot f = e\cdot g$ in $\Nom(M)$.
  Given another cocone, i.e.~an equivariant map $h\colon  Y\to H$ with $h\cdot f
  = h\cdot g$, let $u\colon E\to UH$ be induced by the coequalizer in
  $\SSet{A}$, i.e.~$u$ is the unique supported map with $u\cdot e = h$. This
  supported map is equivariant, because for all $m\in M$ and $x\in E$, we
  straightforwardly have
  \begin{align*}
    m\cdot u(x)
    &= m\cdot u(e(r(x)))
    = m\cdot h(r(x))
    = h(m\cdot r(x))
      \\
    &= u(e(m\cdot r(x)))
      \overset{\text{\eqref{eqEdef}}}{=}
      u(m\cdot e(r(x)))
      = u(m\cdot x).
  \end{align*}
  Hence, $u\colon (E,\cdot)\to H$ is the desired cocone morphism in $\Nom(M)$.
  Uniqueness of $u$ is clear because $U\colon \Nom(M)\to \SSet{A}$ is faithful.
  \qed
\end{proofappendix}
\begin{figure}%
  \begin{minipage}[b]{.48\textwidth}%
    \hspace{-5mm}%
  \begin{tikzpicture}[outer sep=0pt,commutative diagrams/every diagram]
    \matrix[matrix of math nodes, name=m, commutative diagrams/every cell,
    column sep=5mm,row sep=0mm,
    supp/.style={outer sep=2pt,minimum width=8mm},
    ampersand replacement=\&] {
      |[alias=Nom]|
      \overbrace{\Nom}^{\Nom(\Perm(\A))}
      \& |[alias=RnNom]|
      \overbrace{\RnNom}^{\Nom(\Fin(\A))}
      \& |[alias=OrdNom]|
      \overbrace{\OrdNom}^{\Nom(\Aut(\Q,<))}
      \\[18mm]
      |[alias=SuppL,supp]|
      \phantom{X\mathclap{\SSet{A}}}
      \&
      |[alias=Supp,supp]|
      \mathclap{~\SSet{A}~\text{\color{lipicsGray} for countably infinite $A$}}
      \&
      |[alias=SuppR,supp]|
      \mathclap{\phantom{\SSet{A}}}
      \\};
    \begin{scope}[on background layer]
      \node[fit=(SuppL) (SuppR),
      draw=lipicsYellow,
      rounded corners=5pt,
      line width=2pt,
      inner sep=-2pt]  {};
    \end{scope}
    \begin{scope}[adjoint/.style={commutative diagrams/.cd, every arrow, every label, bend left=10,shift left=2}]
      \path[adjoint] (Supp.north -| OrdNom) to node[sloped,name=FQ] {\(\Monad[\Aut(\Q,<)](-)\)} (OrdNom.south) ;
      \path[adjoint] (OrdNom.south) to node[name=UQ] {\(U\)} (Supp.north -| OrdNom) ;
      \path[adjoint] (Supp.north -| RnNom) to node[sloped,name=FR] {\(\Monad[\Fin(\A)](-)\)} (RnNom.south) ;
      \path[adjoint] (RnNom.south) to node[name=UR] {\(U\)} (Supp.north -| RnNom) ;
      \path[adjoint] (Supp.north -| Nom) to node[sloped,name=M] {\(\Monad[\Perm(\A)](-)\)} (Nom.south) ;
      \path[adjoint] (Nom.south) to node[name=U] {\(U\)} (Supp.north -| Nom) ;
    \end{scope}
    \foreach \leftadjoint/\rightadjoint in {M/U,FQ/UQ,FR/UR} {
      \draw[draw=none] (\leftadjoint) -- node[sloped] {\(\dashv\)} (\rightadjoint);
    }
  \end{tikzpicture}
  \caption{Monadic adjunctions (Ex.~\ref{exMonadicity})}
  \label{figMonadicityMSet}
  \end{minipage}%
  \begin{minipage}[b]{.52\textwidth}\centering%
  \begin{tikzpicture}[commutative diagrams/every diagram]
    \matrix[matrix of math nodes, name=m, commutative diagrams/every cell,
            column sep=25mm,row sep=16mm] {
      |[alias=Nom]|
      \Nom(\Perm(\A))
      & |[alias=Presheaf]|
      {\Presheaf}
      \\
      |[alias=Supp]|
      \SSet{\A}
      & |[alias=Indexed]| {\Indexed}
      \\};
    \begin{scope}[adjoint/.style={commutative diagrams/.cd,
        every arrow,
        every label,
        bend left=10,
        shift left=2,
        preaction = {decorate,draw=white, line width=5pt,-,line cap=round},
      },
      every node/.append style={
        fill=white,
        rounded corners=2pt,
        inner sep=1pt,
        outer sep=1pt,
      },
      ]
      \begin{scope}[adjoint/.append style={shift left=-1,bend left=8},
        nomshift/.style={xshift=-4mm,yshift=-1mm},
        idxshift/.style={xshift=4mm,yshift=1mm},
        ]
        \path[adjoint] ([idxshift]Indexed.north west) to node[name=FPNom] {} node[sloped,below] {\(F\cdot \Sigma\)} ([nomshift]Nom.south east) ;
        \path[adjoint] ([nomshift]Nom.south east) to node[name=UPNom] {} node[sloped,above] {\(D\cdot U\)} ([idxshift]Indexed.north west) ;
      \end{scope}
      \begin{scope}[adjoint/.append style={
          shift left=-1,
          bend left=8,
          shorten <= 0pt,
          shorten >= 0pt,
        },]
        \path[adjoint] (Nom.east) to node[name=I] {\(I\)} (Presheaf.west);
        \path[adjoint] (Presheaf.west) to node[name=Istar] {\(I^*\)} (Nom.east) ;
        \path[adjoint] (Supp.east) to node[name=D] {\(D\)} (Indexed.west);
        \path[adjoint] (Indexed.west) to node[name=Sum] {\(\Sigma\)} (Supp.east) ;
      \end{scope}
      \path[adjoint,inner sep=2pt] (Supp.north) to node[sloped,name=M] {\(FX=\Monad[\Perm(\A)]X\)} (Nom.south) ;
      \path[adjoint] (Nom.south) to node[name=U] {\(U\)} (Supp.north) ;
      \path[adjoint] (Indexed.north) to node[name=FT] {\(F^\Idx\)} (Presheaf.south) ;
      \path[adjoint] (Presheaf.south) to node[name=UT] {\(U^\Idx\)} (Indexed.north) ;
    \end{scope}
    \foreach \leftadjoint/\rightadjoint in {Istar/I,M/U,Sum/D,FT/UT,FPNom/UPNom} {
      \draw[draw=none] (\leftadjoint) -- node[sloped] {\(\dashv\)} (\rightadjoint);
    }
    \begin{scope}[
      on background layer,
      area/.style={
        draw=lipicsYellow,
        rounded corners=5pt,
        line width=2pt,
        inner xsep=-1pt,
        inner ysep=2mm,
        yshift=2mm,
        minimum width=25mm,
      }
      ]
      \coordinate (topmid) at ($ (Nom.north) !.5! (Presheaf.north) $);
      \coordinate (botmid) at ($ (Supp.south) !.5! (Indexed.south) $);
      \node[fit=(Nom.north) (Nom) (Supp),area] (NomSupp) {};
      \node[fit=(Presheaf.north) (Presheaf) (Indexed),area] (PreshIndex) {};
      \begin{scope}[every node/.append style={
          anchor=north,
          font=\sffamily\bfseries,
          color=black!80,
          inner ysep=5pt,
        }]
        \node at (NomSupp.north) {Monadic};
        \node at (PreshIndex.north) {Monadic};
      \end{scope}
    \end{scope}
  \end{tikzpicture}%
    \caption{Relation to presheaves for $M:= \Perm(\A)$}
    \label{figDiaPresheaf}
  \end{minipage}%
\end{figure}

\begin{example} \label{exMonadicity}
  The following categories are monadic over $\Supp{A}$, for $A$ being countably
  infinite. The monadic adjunctions are visualized in
  \autoref{figMonadicityMSet}, and the monads listed below.
  \begin{enumerate}[beginpenalty=99,topsep=0pt]
  \item The category $\Nom$ of nominal sets (with equality symmetry) is monadic
    over $\Supp{\A}$. The operations on a generator $x$ in the
    corresponding theory are injective maps $\s{}(x)\monoto \A$:
    \[
      \Monad[\Perm(\A)]X = \set[\big]{
        (\pi,x)
        \mid x\in X, \pi\colon \s{X}(x)\monoto \A
      }
    \]
  \item The category $\RnNom$ of nominal renaming sets is monadic
    over $\Supp{\A}$. The operations on a generator $x$ are arbitrary maps
    $\s{}(x)\to \A$ (not necessarily injective):
    \[
      \Monad[\Fin(\A)]X = \set[\big]{
        (\pi,x)
        \mid x\in X, \pi\colon \s{X}(x)\to \A
      }
    \]
  \item The category $\OrdNom$ of nominal sets for the total order symmetry is
    monadic over $\Supp{\Q}$ (using $\Q\cong A$). The operations on $x$ are
    monotone injective maps $\s{}(x)\monoto \Q$:
    \[
      \Monad[\Aut(\Q,<)]X =
      \big\{
      (\pi,x)
      \mid x\in X, \pi\colon \s{X}(x)\monoto \Q
      ~\forall q,p\in \s{X}(x):
      q < p
      \Rightarrow
      \pi(q) < \pi(p)
      \big\}
    \]

  \end{enumerate}
\end{example}
As a direct application of the monadicity, we can characterize orbit-finite
nominal sets and finitely presentable objects in $\Nom(M)$ using a general
result about algebraic categories~\cite[Thm.~3.7]{amsw19b}:
\begin{example}
  A nominal $M$-set $X$ is finitely presentable iff it can be described by a finite
  supported set $G$ of \emph{generators} and a finite subset
  \(
    E\subseteq (\Monad G) \times (\Monad G)
  \) of \emph{equations}.
  This characterization means that
  given such finite $G$ and $E$, we obtain projection maps
  \[
    \begin{tikzcd}
      E
      \arrow[shift left=1]{r}[overlay]{\ell}
      \arrow[shift right=1]{r}[overlay,swap]{r}
      & \Monad G
    \end{tikzcd}
    \qquad\text{in }\Supp{A}
    \qquad
    \Longleftrightarrow
    \qquad
    \begin{tikzcd}
      \Monad E
      \arrow[shift left=1]{r}[overlay]{\bar \ell}
      \arrow[shift right=1]{r}[overlay,swap]{\bar r}
      & \Monad G
    \end{tikzcd}
    \qquad\text{in }\Nom(M)
  \]
  and their coequalizer in $\Nom(M)$ is $X$.

  For $M:= \Perm(\A)$, 1.~every
  orbit-finite nominal set can be described by such finite supported sets $G$
  and $E$, and 2.~any such finite system of equations $E$ on $G$ presents an
  orbit-finite nominal set.
  For example, the nominal set of unordered pairs of atoms
  can be described by one generator $g$
  encoded as a supported set
  \(
    G=\set{g},
    \s{G}(g) := \set{a,b}
    ~(\text{for fixed }a,b\in \A)
  \)
  and one equation
  \(
    E:=\set[\big]{(a\,b)\cdot g = \id\cdot g}.
  \)
  Here, we use intuitive notation to denote
  \[
    E:=\set[\big]{\big(\,([(a\,b)]_{\set{a,b}},g),~([\id]_{\set{a,b}},g)\,\big)}
    \subseteq (\Monad G)\times(\Monad G).
  \]
  With the projections $\ell,r\colon E\to \Monad G$ (in $\Supp{A}$) and
  their extensions $\bar\ell,\bar r$ to $\Nom$, we have the unordered pairs as a
  coequalizer diagram in $\Nom$:
  \[
    \begin{tikzcd}
        \Monad E
        \arrow[shift left=1]{r}{\bar \ell}
        \arrow[shift right=1]{r}[swap]{\bar r}
        & \Monad G \arrow[->>]{r}
        & \set[\big]{\{c,d\}\mid c,d\in \A, c\neq d}.
    \end{tikzcd}
  \]
\end{example}

\begin{remark}[Relation to Presheaves]
  \label{remPresheaf}
The category of supported sets $\SSet{\A}$ nicely fits into an existing diagram of
Kurz, Petrisan, and Velebil~\cite{KPVAlgTheoriesNom} relating nominal sets
$\Nom$ (for the equality symmetry $\Perm(\A)$) with two presheaf categories:
\begin{enumerate}
\item $[\Idx,\Set]$ is the category of functors $P\colon \Idx\to\Set$
  (\emph{sets in context}), where the objects of $\Idx$ are finite subsets of
  $\A$ (i.e.~$|\Idx|=\Powf \A$) and the morphisms are injective maps (i.e.~$\Idx
  \neq (\Powf\A,\subseteq)$).
\item $[|\Idx|,\Set]$ is the category of functors $P\colon |\Idx|\to
  \Set$, from the set $\Powf(\A)$ to $\Set$, i.e.~$P$ is a
  $\Powf(\A)$\hyp{}indexed family of sets.
\end{enumerate}

\autoref{figDiaPresheaf} shows the result when extending the diagram of
Kurz~\etal~\cite{KPVAlgTheoriesNom}  by $\Supp{\A}$.
Let us go through the functors and adjunctions relating the categories:
\begin{enumerate}
\item The left adjoint $\Sigma\colon \Indexed\to \Supp{\A}$
  sends a family $X\colon \Powf(\A)\to \Set$ to the coproduct
  $\Sigma(X) = \coprod_{S\in \Powf(\A)} X(S)$,
  where the component for $S\in \Powf(\A)$ has support $S$.

\item The right-adjoint $D\colon \Supp{\A}\to \Indexed$ forms down-sets: for a
  supported set $X$, the family $DX\colon \Powf\A\to \Set$ is given by
  \(
    DX(S) = \{x\in X\mid \s{X}(x)\subseteq S\}.
  \)
\item Since adjunctions compose, we have the adjunction
  $F\cdot \Sigma \dashv D\cdot U$, which is \emph{not
    monadic}~\cite{KPVAlgTheoriesNom}, so the
  monad $DUFR$ does not have $\Nom$ as its Eilenberg-Moore category.
\item Instead, $\Presheaf = \EM(DUFR)$~\cite{KPVAlgTheoriesNom}, and $F^\Idx \vdash U^\Idx$ is the
  corresponding adjunction.
\item The induced comparison functor from $\Nom$ is itself adjoint: $\Nom$ is
  (equivalent to) the full reflective subcategory of pullback preserving
  functors (cf.~\cite{GadducciEA2006,GabbayP99}).\twnote[inline]{}
\end{enumerate}
\end{remark}
\begin{proofappendix}[Details for]{remPresheaf}
  The full definitions on the functors mentioned in \autoref{remPresheaf} are as
  follows:
  \begin{itemize}
  \item The left adjoint $\Sigma\colon \Indexed\to \Supp{\A}$
    sends $X\colon \Powf(\A)\to \Set$ to the supported set
    \[
      \Sigma(X) = \coprod_{S\in \Powf(\A)} X(S)
      \qquad
      \s{\Sigma(X)}(\inj_S(x)) = S
    \]
    and a natural transformation $f\colon X\to Y$ to the supported map
    \[
    \Sigma(f)\colon \Sigma X\to \Sigma Y
    \quad
    \Sigma(f)(\inj_S(x)) = \inj_S(f(x)).
    \]
  \item The right-adjoint $D\colon \Supp{\A}\to \Indexed$ forms down-sets: for a
    supported set $X$, the family $DX\colon \Powf\A\to \Set$ is given by
    \[
    DX(S) = \{x\in X\mid \s{X}(x)\subseteq S\}.
    \]
    For a supported map $f\colon X\to Y$ the natural transformation $Df\colon
    DX\to DY$ is given by
    \[
      (Df)_S\colon DX(S)\to DY(S)\quad
      (Df)_S(x) = f(x).
    \]

  \item The composition $D\cdot U\colon \Nom\to \Indexed$ sends a nominal set
    $X$ to a family $DUX$ where for $S\in\Powf(X)$, $DUX(S) \subseteq X$
    contains all elements supported by $S$; this is precisely the functor
    mentioned by Kurz \etal~\cite{KPVAlgTheoriesNom}.
    \hfill\qed\qedhere
  \end{itemize}
\end{proofappendix}
Note that $\Supp{\A}$ is the only category in \autoref{figDiaPresheaf} that is
not a topos, and therefore not a presheaf category.

\begin{remark}[Relation to named sets]
  Another notion to reason about names are \emph{named sets}~\cite{FMP02},
  which have been defined in slightly different ways over the years. For a fixed
  countably infinite set of names $\A=\set{v_1,v_2,\ldots}$ the definitions are as follows:
  \begin{itemize}
  \item In the definition by Ferrari, Montanari, Pistore~\cite[Def.~2]{FMP02},
  every element $x$ of a named set $X$ does not have an explicit support, but
  is only associated with a natural number $n\in \N$, and so its support is
  implicitly considered to be $\set{v_1,\ldots,v_{n}}\subseteq \A$. Moreover,
  every element is equipped with a subgroup of $\Perm(n)$ that describes how the atoms
  $v_1,\ldots,v_n$ in the support may be permuted. In later works, this natural
  number $n\in \N$ denoting only the size was replaced with a set of atoms:

  \item Montanari and Pistore~\cite{MontanariP05} define a named set to be
  a set $X$ together with a map $n_X\colon X\to \Pow (\A)$ (without any further
  information about permutations). A \emph{named function} $m\colon (X,n_X)\to
  (Y,n_Y)$ is a map on the sets $m\colon X\to Y$ and for each $x\in X$ an
  injective map $m[x]\colon n_Y(f(x)) \monoto n_X(x)$. A named set $(X,n_X)$ is
  called \emph{finitely named} if $n_X(x)$ is finite for every $x\in X$.

  The category of all named sets and named functions is quite different from $\Supp{\A}$. For instance,
  the finitely named set $(\A,n_{\A})$ with $n_{\A}(a) = \set{a}$ has infinitely many automorphisms $(\A,n_{\A}) \to (\A,n_{\A})$, whereas both in $\Nom(\Perm(\A))$ and in $\Supp{\A}$, there is only one morphism $\A\to \A$, the identity.
  Without the finiteness restriction, infinite limits in the category of named sets are
  \Set-like and not \Nom-like. More precisely, the infinite product
  $\prod_{n\in \N} (\A,n_{\A})$ contains streams making use of infinitely many names.

  \item Gaducci, Miculan, and Montanari~\cite{GadducciEA2006} restrict to
  \emph{finitely} named sets and additionally equip every element $x\in X$ of a named set $(X,n_X)$ with a subgroup
  $G_X(x)$ of the group $\Perm(n_X(x))$ of all permutations on the names
  $n_X(x)$. Here, a named function $f\colon (X,n_X,G_X)\to (Y,n_Y, G_Y)$, is a map $f\colon X\to Y$ and additionally for each $x\in X$
  a non-empty set of injections $f_x \subseteq (n_Y(f(x)) \monoto n_X(x))$ satisfying
  an additional coherence condition involving $G_X$. Details are not relevant
  here, because the resulting category $\NSet$ was shown to be equivalent to
  $\Nom(\Perm(\A))$ in the same work~\cite[Prop.~29]{GadducciEA2006}. This has
  a couple of immediate implications regarding the relationship to supported
  sets:
  \begin{enumerate}
  \item $\NSet$ is monadic over $\Supp{\A}$.
  \item The right-adjoint of this monadic adjunction is a
  faithful functor $\Supp{\A}\to \NSet$, which is not full.
  \end{enumerate}
  \end{itemize}
\end{remark}

Having discussed categorical properties of supported sets and its relation to
other name-aware notions, we now continue to see how name binding can be accomplished.

\section{Name Abstraction and de Bruijn indices}
\label{secAbstraction}
In $\lambda$-calculus, the computational steps ($\beta$-reduction) essentially
consist of a substitution rule
\(
  (\lambda x.\,T)\, P \longrightarrow_\beta T[x := P]
\)
which requires some bound variable names in $T$ to be
sufficiently fresh. Thus, it might be necessary to rename those bound variables
($\rightarrow_\alpha$) in $T$ before a $\beta$ step can be performed. But even
in such a renaming step $\rightarrow_\alpha$,
a similar side-condition needs taken care of, because otherwise, the reduction would lead
to wrong results.

Thus, there are several approaches to define $\lambda$-expressions directly
modulo $\alpha$-equivalence, making substitution total and $\beta$-reduction
always applicable to $\lambda$-expressions containing a reducible expression
(redex).
In 1972, de Bruijn~\cite{debruijn1972} invented a technique of replacing the
variable names with an index (the \emph{de Bruijn index}) that counts the number
binders between the variable and its corresponding binder. In 1999, Gabbay and
Pitts~\cite{GabbayP99} presented another way to define
$\lambda$-expressions directly as $\alpha$-equivalence classes,
in which
$\lambda$-abstraction is a functor on \emph{nominal sets} (called \emph{FM-sets}
back then). From now on, we stick to their setting of equality symmetry:
\begin{assumption}
  \label{assCountable}
  For the rest of the paper, fix $A:=\A$ and $M := \Perm(\A)$.
  Hence, we may assume a bijection $\rho\colon \N\to \A$ and we
  simply write $\Nom$ for the $M$-nominal sets.
\end{assumption}

For the name abstraction functor, the notion of $\alpha$-equivalence is first
defined for arbitrary nominal sets, which is then used in the definition of the
abstraction functor:

\begin{notheorembrackets}
  \begin{definition}[{\cite{GabbayP99}}]
    \label{defAbs}
    For $X\in \Nom$, the equivalence relation $\sim_\alpha$
    on $\A\times X$ is defined by
    \[
      (a,x) \sim_\alpha (b,y)
      ~~:\Leftrightarrow~~
      \exists c \fresh (a,b,x,y)\colon (c\, a)\cdot x = (c\,b)\cdot y
    \]
    The \emph{abstraction functor} $\Abs X\colon \Nom\to \Nom$
    is given by
    \(
    \Abs X = (A\times X)/\mathord{\sim_\alpha}
    \),
    where $\abs{a}{x}$ denotes the equivalence class of $(a,x)\in \A\times X$.
  \end{definition}
\end{notheorembrackets}
In the equivalence class $\abs{a}x$, $a$ disappears from the support:
\(
\supp{\Abs X}(\abs{a}{x}) = \supp{X}(x) \setminus \{a\}.
\)

\begin{notheorembrackets}
  \begin{example}[{\cite{GabbayP99}}] \label{exLambdaFunctor}
    The initial algebra of the functor
    \(
    \Lambda X = \A + \Abs X + X\times X
    \)
    is carried by the nominal set of $\lambda$-expressions modulo
    $\alpha$-equivalence. The first summand $\A$ describes variables $a\in \A$,
    the second summand $\Abs X$ describes $\lambda$-abstractions $\lambda x. T$,
    where $x\in \A$ and $T$ is a $\lambda$-expression, and the third summand
    $X\times X$ describes the application $T\,S$ of one $\lambda$-expression to one
    other.
  \end{example}
\end{notheorembrackets}

In the present paper, we define an abstraction functor on supported sets
for which we require the set of atoms to be a countably
infinite set. This functor in fact has a lifting to nominal sets and the
lifting is (naturally isomorphic to) the above abstraction functor $\Abs\colon
\Nom\to\Nom $ on nominal sets.

The definition of $\Abs$ makes use of the $\Perm(\A)$-action to capture
$\alpha$-equivalence. When introducing abstraction as a functor directly on supported sets,
we do not have renaming available, and so we use de Bruijn indices via the bijection $\rho\colon
\N\to \A$ (\autoref{assCountable}).
\begin{definition} \label{defB}
  The \emph{de Bruijn functor} $\B\colon \Supp{\A}\to\Supp{\A}$ sends a supported set $X$ to
  the same set $\B X = X$ but with a different support function. To distinguish
  elements of $X$ and $\B X$, we write $\bind x\in \B X$ for $x\in X$ (i.e.~$\lambda$
  is a nameless binder). The
  support on $\B X$ is
  \begin{align}
    \s{\B X}\colon \B X\to \Powf(\A)
    \qquad
    \s{\B X}(\bind x) := \{\rho (k) \mid \rho(k+1)\in \s{X}(x), k\in \N\}.
    \label{defSB}
  \end{align}
  A supported map $f\colon X\to Y$ is sent to
  the same map
  \(
    \B f\colon \B X\to \B Y
    \),
    \(
    \B f(\bind x) = \bind f(y).
  \)
\end{definition}
\begin{figure}
  \centering
  \(
  \lambda x. x (\lambda x.x)(\lambda y.xy)
  \)
  ~~
  \begin{tikzcd}[arrow style=tikz,>={Straight Barb[scale=0.8]},]
    {}
    \arrow[<->,line width=1.4pt,draw=lipicsYellow,
      ]{r}{\textsf{Natural Isomorphism $\phi$}}[below]{{\textsf{from \autoref{absLifts}}}}
    &[3cm]
    {}
  \end{tikzcd}
  ~~
  \begin{tikzpicture}[baseline=(bx.base),
    every node/.append style={
      anchor=base west,
      alias = mylastnode,
      inner xsep=2pt,
    },
    mapper/.style={
      draw=blue,
      rounded corners=2pt,
      line width=1pt,
      draw=lipicsYellow,
      ->,
    },
    ]
    \node (bx) {$\bind$};
    \node[] (x1) at (mylastnode.base east) {$\varrho(0)$};
    \node[] (bz) at (mylastnode.base east) {${(}\bind$};
    \node[] (z1) at (mylastnode.base east) {$\varrho(0){)}$};
    \node[] (by) at (mylastnode.base east) {${(}\bind$};
    \node[] (x2) at (mylastnode.base east) {$\varrho(1)$};
    \node[] (y1) at (mylastnode.base east) {$\varrho(0){)}$};
    \foreach  \src/\targ in {x1/bx,z1/bz,y1/by} {
      \path[mapper,overlay] (\src.south) --
      ([yshift=-4mm]\src.south)
      -| (\targ.south);
    }
    \foreach  \src/\targ in {x2/bx} {
      \path[mapper] (\src.north) --
      ([yshift=4mm]\src.north)
      -| (\targ.north);
    }
  \end{tikzpicture}
  \caption{Visualization of de Bruijn indices}
  \label{figVisBruijn}
\end{figure}
The definition of $\s{\B X}$ captures the idea of de Bruijn indices:
we can think of the notation $\bind x$ as a lambda abstraction of nameless
binder $\bind$ of a lambda term $x$ (visualized in \autoref{figVisBruijn}):
\begin{itemize}[topsep=3pt]
\item The variables referring to $\bind$ have the de Bruijn index of 0 (at the
  level of $x$), because there is no other binder between $x$ and \textqt{$\bind$}. Since
  $\rho(0)$ is bound, $\s{\B X}(x)$ does not depend on whether $\rho(0)\in
  \s{X}(x)$.

\item All other variables $\rho(k+1)\in \s{X}(x)$ refer to variables
  that are free in $\bind x$ and so refer to binders \textqt{more above} than $\bind
  x$. A variable $\rho(k) \in \s{\B X}(\bind x)$ refers
  to the same binder as the variable $\rho(k+1)\in \s{X}(x)$ under \textqt{$\bind$},
  because the latter is one level further down.
\end{itemize}

\begin{lemma}
  \label{Bfunctorial}
  $\B\colon \Supp{\A}\to\Supp{\A}$ is a functor.
\end{lemma}
\begin{proofappendix}{Bfunctorial}
  For every supported set $X$ (with $\s{X}$), $\B X$ (with $\s{\B X}$) is a
  supported set by definition. A supported map $f\colon X\to Y$ is sent by $\B$
  to the map
  \[
    \B f\colon \B X\to \B Y
    \qquad
    \B f(\bind x) = \bind f(y).
  \]
  This is again a supported map because
  \begin{align*}
      \s{\B Y}(\B f(\bind x))
    &= \s{\B Y}(\bind f(x))
    \\ &= \set{\rho(k) \mid \rho(k+1) \in \s{Y}(f(x)),k\in \N}
      \tag*{\eqref{defSB}}
    \\ &\subseteq \set{\rho(k) \mid \rho(k+1) \in \s{X}(x),k\in \N}
    \\ &= \s{\B X}(\bind x)
         \tag*{\eqref{defSB}}
  \end{align*}
  Since $\B f$ just has $f$ as its underlying map, $\B$ preserves identities and
  compositions of supported maps.
\end{proofappendix}
\noindent
We use a slightly generalized notion of a \emph{lifting} of the functor $\B$ to
nominal sets:

\smallskip
\noindent
\begin{minipage}[C]{.77\textwidth}
\begin{definition} \label{defLift}
  For a monad $T$ on a category $\C$ and a functor $H\colon \C\to\C$, a
  functor $G\colon \EM(T)\to\EM(T)$ is called a
  \emph{lifting of $H$} if $HU$ and $UG$ are naturally isomorphic functors.
  We say that a lifting is \emph{strict} if $HU = UG$.
\end{definition}
\end{minipage}%
\begin{minipage}[C]{.23\textwidth}
  \hfill
  \begin{tikzcd}[row sep=5mm,column sep=5mm]
    \EM(T)
    \arrow{r}{G}
    \arrow{d}[swap]{U}[alias=u1]{}
    & \EM(T)
    \arrow{d}{U}[swap,alias=u2]{}
    \\
    \C \arrow{r}{H}
    & \C
  \end{tikzcd}%
  \!\!%
\end{minipage}

\begin{remark}
  Usually, not just a natural
  isomorphism but identity is required. Strict liftings $G\colon \EM(T)\to\EM(T)$ are in
  one-to-one correspondence to distributive laws $TH\to HT$~\cite{johnstone75Lifting}.
\end{remark}

\noindent
This generalization to natural isomorphisms is sound, because they induce strict liftings:
\begin{lemma} \label{liftUpToIso}
  For every natural isomorphism $\phi\colon HU\to UG$,
  there is a unique strict lifting
  \(
    \bar H\colon \EM(T)\to\EM(T)
  \)
  such that $\phi\colon H\to G$ is a natural isomorphism in $\EM(T)$.
\end{lemma}
This means that for $(C,\gamma)\in \EM(T)$, $\phi_{(C,\gamma)}$ is a
$T$-algebra isomorphism $\bar H(C,\gamma)\to G(C,\gamma)$.
\begin{proofappendix}{liftUpToIso}
  In the following, we denote the algebra structure of an
  Eilenberg\hyp{}Moore\hyp{}algebra $(C,\alpha)$ by
  \[
    \struct(C,\alpha) := \alpha.
  \]
  Since the carrier of an Eilenberg-Moore-algebra is given by $U(C,\alpha) = C$,
  we can write every $T$-algebra $X := (C,\alpha)$ as a morphism
  \[
    T U X \xrightarrow{\struct(X)} U X
  \]
  without needing to \textqt{unpack} the tuple $(C,\alpha)$.
  In order to be a strict lifting, the functor $\bar H\colon \EM(T)\to\EM(T)$
  needs to send a $T$-algebra $X := (C,\alpha)$ to a $T$-algebra on $HC$, so it only
  remains to define $\struct(\bar H X)$ as the composition:
  \[
    \begin{tikzcd}
      T H U X
      \arrow[dashed]{d}[swap]{\struct(\bar HX)}{:=}
      \arrow{r}{T\phi_X}
      & T U G X
      \arrow{d}{\struct(GX)}
      \\
      H U X
      \arrow[<-]{r}{\phi_X^{-1}}
      & U G X
    \end{tikzcd}
  \]
  Hence, we have that $\phi_X$ is a homomorphism of $T$-algebras by definition and that
  $\struct(\bar HUX)$ is the unique morphism with this property.
  The unit $\eta$ of the monad $T$ is preserved, because $\struct(\bar HX)\cdot
  \eta_{HUX} = \id_{HUX}$.
  \[
    \begin{tikzcd}
      H U X
      \arrow{d}[swap]{\eta_{HUX}}
      \arrow{r}{\phi_X}
      \descto{dr}{Naturality}
      & U G X
      \arrow{d}{\eta_{UGX}}
      \arrow[shiftarr={xshift=14mm}]{dd}{\id_{UGX}}
      \\
      T H U X
      \arrow{d}[swap]{\struct(\bar HX)}
      \arrow{r}{T\phi_X}
      \descto{dr}{Def.}
      & T U G X
      \arrow{d}{\struct(GX)}
      \\
      H U X
      \arrow[<-]{r}{\phi_X^{-1}}
      & U G X
    \end{tikzcd}
  \]
  With the same successive application of the definition of $\struct(\bar HX)$,
  it preserves the multiplication $\mu$ of $T$:
  \[
    \begin{tikzcd}[bend angle=5]
      TTHUX
      \arrow{rrr}{T\struct(\bar H X)}
      \arrow{ddd}{\mu_{HUX}}
      \arrow{dr}{TT\phi_X}
      &
      & & THUX
      \arrow[->,bend left]{dl}{T\phi_X}
      \arrow[<-,bend right]{dl}[swap]{T\phi_X^{-1}}
      \arrow{ddd}[swap,pos=0.65]{\struct(\bar HX)}
      \\
      \descto[yshift=3mm]{dr}{Naturality}
      & TTUGX
      \arrow{d}[swap]{\mu_{UGX}}
      \arrow{r}[yshift=2mm]{T\struct(G X)}
      \descto{dr}{Axiom}
      & TUGX
      \arrow{d}{\struct(G X)}
      \\
      & TUGX
      \arrow{r}[swap,yshift=-1mm]{\struct(G X)}
      & UGX
      \\
      THUX
      \arrow[->]{ur}{T\phi_X}
      \arrow{rrr}{\struct(\bar HX)}
      & & & HUX
      \arrow[<-]{ul}{\phi_X^{-1}}
    \end{tikzcd}
  \]
  For the functoriality of $\bar H$, it is easy to see that every $T$-algebra
  homomorphism $f\colon X\to Y$ is sent to a homorphism again since $G$ is
  functorial:
  \[
    \begin{tikzcd}
      THUX
      \arrow[shiftarr={yshift=5mm}]{rrr}{\struct(\bar HX)}
      \arrow{r}{T\phi_X}
      \arrow{d}[swap]{THUf}
      \descto{dr}{Naturality}
      & TUGX
      \arrow{r}{\struct(GX)}
      \arrow{d}{TUGf}
      &[4mm] UGX
      \arrow{r}{\phi_X^{-1}}
      \arrow{d}[swap]{UGf}
      \descto{dr}{Naturality}
      & HUX
      \arrow{d}{HUf}
      \\
      THUY
      \arrow[shiftarr={yshift=-5mm}]{rrr}{\struct(\bar HY)}
      \arrow{r}{T\phi_Y}
      & TUGY
      \arrow{r}{\struct(GY)}
      &[4mm] UGY
      \arrow{r}{\phi_Y^{-1}}
      & HUY
    \end{tikzcd}
  \]
  Finaly, preservation of identities and composition is immediate.
\end{proofappendix}

\smallskip\noindent
\begin{minipage}{.73\textwidth}
\begin{theorem}
  \label{absLifts}
  The abstraction functor $\Abs\colon \Nom\to \Nom$ is a lifting of 
  the de Bruijn functor $\B\colon \Supp{\A}\to \Supp{\A}$. That is, there is a
  natural isomorphism $\phi\colon \B U \longrightarrow U\Abs$.
\end{theorem}
\end{minipage}%
\begin{minipage}{.27\textwidth}
  \hfill
  \begin{tikzcd}[row sep=5mm,column sep=5mm]
    \Nom
    \arrow{r}{\Abs}
    \arrow{d}[swap]{U}[alias=u1]{}
    & \Nom
    \arrow{d}{U}[swap,alias=u2]{}
    \\
    \Supp{\A} \arrow{r}{\B}
    & \Supp{\A}
  \end{tikzcd}%
  \!\!%
\end{minipage}

\begin{proofappendix}{absLifts}
  The natural isomorphism $\phi\colon \B U \longrightarrow U\Abs$
  is defined by
  \begin{align*}
    \phi_X&\colon \B U X\longrightarrow U\Abs X
    \\
    \phi_X&(\bind x) = \sigma_{\maxidx(x)}^{-1}\cdot \abs{\rho(0)} x
  \end{align*}
  where $\maxidx(x)\in \N$ and $\sigma_m\in \Perm(\A)$ ($m\in \N$) are given by:
  \begin{align*}
    \maxidx(x) &= 1 + \max\set{n\in \N\mid \rho(n) \in \s{}(x)}
    \\
    \sigma_m(\rho(\ell)) &= \big(\rho(0)\cdots \rho(m)\big)
                           = \begin{cases}
                             \rho(\ell+1)\!\!\!
                             &\text{if }\ell < m
                             \\
                             \rho(0)
                             &\text{if }\ell = m
                             \\
                             \rho(\ell)&\text{else.}
                           \end{cases}
    \\
  \end{align*}

  We verify that $\phi$ is indeed a natural isomorphism between $\B U$ and
  $U\Abs$ by showing that every $\phi_X$ is a support"=reflecting, bijective
  map (\autoref{propSuppIso}), and that it is natural in $X$:
  \begin{description}[itemsep=2pt,leftmargin=0pt]
    \item[Support-reflecting:]
      For $\bind x \in \B U X$, let $m := \maxidx(x)$:
      \begin{align*}
        &\mathbin{\phantom{=}}\s{U\Abs X}(\phi_X(\bind x)) \\
        &= \s{U\Abs X}(\sigma_m^{-1}\cdot \abs{\rho(0)}x) \\
        &= \sigma_m^{-1}\cdot \supp{\Abs X}(\abs{\rho(0)}x)
          \tag{\autoref{suppRenameIncl}}
        \\
        &= \sigma_m^{-1}\cdot (\supp{X}(x)\setminus{\rho(0)})
        \\ &= \sigma_m^{-1}\cdot \set{\rho(k)\mid \rho(k) \in \supp{X}(x), k \neq 0}
        \\ &= \set{\sigma_m^{-1}(\rho(k))\mid \rho(k) \in \supp{X}(x), k \neq 0}
        \\ &= \set{\rho(k-1)\mid \rho(k) \in \supp{X}(x), k \neq 0}
             \tag{Def.~$\sigma_m$}
        \\ &= \set{\rho(k)\mid \rho(k+1) \in \supp{X}(x), k \in \N}
        \\ &= \set{\rho(k)\mid \rho(k+1) \in \s{UX}(x), k \in \N}
        \\ &= \s{\B UX}(\bind x) \tag{by \eqref{defSB}}
      \end{align*}
      So $\phi_X$ is indeed support-reflecting.
    \item[Surjective:]
      For the surjectivity of $\phi_X\colon \B U X\to U\Abs X$, consider
      $\abs{\rho(k)}y\in U\Abs X$. With
      \[
        m:= \max\{\maxidx(y), k\} + 1
      \]
      and
      \[
        x := \big(\big(\rho(0)~\rho(k+1)\big)
        \cdot 
        \sigma_{m}
        \cdot y
        \big),
      \]
      we will that $\phi_X(\bind x) = \abs{\rho(k)}y$.

      For the definition of $\phi_X$, let us first investigate $\maxidx(x)$:
      \begin{align*}
        \supp{}(\sigma_m\cdot y) &= \set{
                                   \rho(\ell+1)\mid \rho(\ell)\in \supp{}(y)}
      \end{align*}
      and moreover
      \begin{align*}
        \qquad%
        \supp{}(x) &\subseteq \set{\rho(0)}\cup\set{
                                   \rho(\ell+1)\mid \rho(\ell)\in \supp{}(y)}
      \end{align*}
      Hence, $\maxidx(x) \le \maxidx(y) < m$. Every $\rho(\ell)$ in the support
      of $\abs{\rho(0)}x$ is in the range $1\le \ell < m$, so
      \[
        \sigma_{\maxidx(x)}^{-1}(\rho(\ell))
        = \rho(\ell) - 1
        = \sigma_m^{-1}(\rho(\ell))
      \]
      and $\sigma_m^{-1}\resteq{S}\sigma_{\maxidx(x)}^{-1}$
      for $S := \supp{}(\abs{\rho(0)}x)$
      (\autoref{defRestEq}). Since $S$ supports $\abs{\rho(0)}x$
      (\autoref{defNominal}), these permutations act identically on
      it, i.e.
      \[
        \sigma^{-1}_{\maxidx(x)}
        \cdot \abs{\rho(0)}x
        =\sigma^{-1}_{m}
        \cdot \abs{\rho(0)}x
      \]
      and we verify:
      \begin{align*}
        \phi_X(\bind x)
        &= \sigma^{-1}_{\maxidx(x)}
        \cdot \abs{\rho(0)}x
        \tag{Def.~$\phi_X$}
        \\ & = \sigma^{-1}_{m}
             \cdot \abs{\rho(0)}x
      \end{align*}
      Note that $\rho(0)\notin \supp{}(\sigma_{m}\cdot y)$, so
      $\rho(k+1) \notin \supp{}(x)$, i.e.~$\rho(k+1)$ is fresh for $x$. Hence,
      \[
        (\rho(0),x) \sim_\alpha
        (\rho(k+1), (\rho(0)~\rho(k+1))\cdot x)
      \]
      and we conclude:
      \allowdisplaybreaks
      \begin{align*}
        \qquad %
        \phi_X(\bind x)
        &= \sigma^{-1}_{m}
          \cdot \abs{\rho(0)}x
        \\
        &= \sigma^{-1}_{m}
          \cdot \abs{\rho(k+1)}\big((\rho(0)~\rho(k+1))\cdot x\big)
        \\
        &= \sigma^{-1}_{m}
          \cdot \abs{\rho(k+1)}\big(
          \sigma_{m}
          \cdot y
          \big)
          \tag{Def.~$x$}
        \\
        &= 
          \abs{\sigma^{-1}_{m}(\rho(k+1))}\big(
          \sigma^{-1}_{m} \cdot \sigma_{m}
          \cdot y \big)
          \tag{Def.~$\Abs$}
        \\
        &= 
          \abs{\sigma^{-1}_{m}(\rho(k+1))}\big(
          y \big)
        \\
        &= 
          \abs{\rho(k)}\big(
          y \big)
          \tag{$k<m$}
      \end{align*}

    \item[Injective:]
      Consider $x,y$ with
      \begin{align*}
        \phi_X(\bind x)
        &= \sigma_{\maxidx(x)}^{-1}\cdot \abs{\rho(0)}x \\
        &= \sigma_{\maxidx(y)}^{-1}\cdot \abs{\rho(0)}y
        = \phi_X(\bind y).
      \end{align*}
      So by assumption already, $\abs{\rho(0)}x$, $\abs{\rho(0)}y\in \Abs X$ are
      in the same orbit. Since we have proven the support-reflectivity of
      $\phi_X$ already, we have
      \begin{align*}
        \s{\B UX}(\bind x))
        &= \supp{\Abs X}(\phi_X(\bind x))
          \\
        &= \supp{\Abs X}(\phi_X(\bind y))
          = \s{\B UX}(\bind y)).
      \end{align*}
      By the definition of $\s{\B UX}$, we have
      \[
        \s{UX}(x) \cup {\rho(0)}
        = \s{UX}(y) \cup {\rho(0)}.
      \]
      Since $\abs{\rho(0)}x$, $\abs{\rho(0)}y$ are in the same orbit,
      $\rho(0)\in \supp{}(x)$ iff $\rho(0)\in \supp{}(y)$. Hence,
      \[
        \s{UX}(x) = \s{UX}(y)
      \]
      and $\maxidx(x) = \maxidx(y)$. So in fact $\abs{\rho(0)}x = \abs{\rho(0)}y$
      and by~\cite[Lemma 4.2]{pittsbook} also $x = y$ and $\bind x=\bind y$ as desired.
    \item[Natural:]
      We need to verify that for every equivariant map $f\colon X\to Y$, the
      diagram
      \[
        \begin{tikzcd}
          \B U X
          \arrow{r}{\phi_X}
          \arrow{d}[swap]{\B U f}
          & U \Abs X
          \arrow{d}{U \Abs f}
          \\
          \B U Y
          \arrow{r}{\phi_Y}
          & U \Abs Y
        \end{tikzcd}
      \]
      commutes. To this end, we verify:
      \begin{align*}
        &\mathbin{\phantom{=}}U\Abs f(\phi_X(\bind x))
          \\
        &= U\Abs f(\sigma_{\maxidx(x)}^{-1}\cdot \abs{\rho(0)}x)
        \\
        &= \sigma_{\maxidx(x)}^{-1}\cdot U\Abs f(\abs{\rho(0)}x)
        \tag{$\Abs f$ equivariant}
        \\
        &= \sigma_{\maxidx(x)}^{-1}\cdot \abs{\rho(0)}(f(x))
      \end{align*}
      Since every $\rho(\ell)$ $\in$ $\supp{}(\abs{\rho(0)}(f(x)))$ is in the range
      \[
        1\le \ell < \maxidx(f(x))\le \maxidx(x)
      \]
      we have
      \[
        \sigma^{-1}_{\maxidx(x)}(\rho(\ell))
        = \rho(\ell - 1)
        = \sigma^{-1}_{\maxidx(f(x))}(\rho(\ell)).
      \]
      Thus, we can conclude:
      \begin{align*}
        &\mathbin{\phantom{=}}U\Abs f(\phi_X(\bind x))
        \\
        &= \sigma_{\maxidx(x)}^{-1}\cdot \abs{\rho(0)}(f(x))
        \\
        &= \sigma_{\maxidx(f(x))}^{-1}\cdot \abs{\rho(0)}(f(x))
        \\
        &= \phi_Y(\bind f(x))\tag{Def.~$\phi_Y$}
        \\
        &= \phi_Y(\B U f(\bind x))
          \tag*{\qedhere}
      \end{align*}
    \end{description}
\end{proofappendix}
Intuitively, $\phi$ translates between the de Bruijn indices and the nominal
abstraction functor: in the term $\bind x\in \B X$, we have $\rho(0)$ implicitly bound,
like it is in $\abs{\rho(0)}(x)\in \Abs X$. However, every $\rho(k+1)$ in $x$
appears as $\rho(k)$ in the support of $\bind x$, so $\phi$
essentially renames $\rho(\ell+1) \mapsto \rho(\ell)$ in $x$ for all $\ell \ge 1$.

\begin{remark}[Abstraction in named sets]
  A de Bruijn-style abstraction functor can also defined directly on $\Nom$~\cite[Def.~3.3]{CianciaM08}.
  The defining property~\eqref{defSB} of abstraction on $\Supp{\A}$ then reappears as a theorem of the support
  function $\supp{}$ in nominal sets~\cite[Thm.~4.1]{CianciaM08}. This is adheres to
  the principle that in $\Supp{\A}$, the support is part of the data, whereas
  in nominal sets, $\supp{}$ is a derived notion.
  Since named sets are equivalent to nominal sets, we have name
  abstraction there, too, and an explicit definition is provided by Ciancia and Montanari~\cite[Sect.~7.1]{CianciaM08}.
\end{remark}

\begin{remark}[Abstraction in presheaves]
  Fiore~\etal~\cite{FiorePlotkinTuri99} also use de Bruijn indices in their
  abstraction functor, but treat support in a different way. They consider
  presheaves $X\in \Set^\F$ where $\F$ is the full subcategory of $\Set$
  containing only natural numbers as objects (considering natural numbers as
  finite cardinals). Thus for every natural number $n\in \N$, the set $X(n)$
  intuitively denotes the elements that make only use of the
  atoms $\rho(0),\ldots,\rho(n-1) \in \A$. For each map $\pi\colon
  n\to m$, the map $X(\pi)\colon X(n)\to X(m)$ renames embedded variables.
  Their name abstraction functor $\delta\colon \Set^\F\to \Set^\F$ sends
  a presheaf $X$ to the presheaf $\delta(X)(n) = X(n+1)$, implicitly binding
  $\rho(0)$, and \textqt{shifting} the role of the other atoms. Despite this
  similarity, it is unclear whether there is a formal categorical  connection
  between $\Supp{A}$ and $\Set^\F$ or between $\B$ and $\delta$.
\end{remark}

With name binding in supported sets, we can now study automata in supported sets
and relate them to nominal automata.

\section{Register Automata to Nominal Automata}
\label{secGenDet}
The well-known powerset construction for automata is an instance of a more
general principle called \emph{generalized determinization}~\cite{SilvaBBR13,bartelsphd} that
\emph{internalizes side-effects} modelled by a monad.
The input of the standard powerset construction is a \emph{non-deterministic} finite
automaton (NFA), which can be understood as a deterministic automaton extended with
non-deterministic branching as a side-effect. With $Q$ as the set of states of the NFA,
the construction returns a deterministic automaton with states $\Powf Q$,
which happens to be precisely the set of \emph{configurations} of the original NFA.
This construction generalizes from the (finite) powerset monad $\Powf$ to arbitrary monads $T\colon \C\to \C$
and from automata to state-based systems modelled via coalgebras for a functor $H\colon \C\to \C$.
We apply this principle to the monad from the
monadicity result in \autoref{secMonadicity}.

A \emph{coalgebra} (for $H\colon \C\to \C$) is an object $Q\in \C$ together with
a morphism $Q\to HQ$. E.g.~an $H$-coalgebra for $H\colon\Set\to\Set$, $HX=2\times X^\Sigma$, is a deterministic automaton, i.e.~a set $Q$ together with a map
\(d\colon Q\to 2\times Q^\Sigma\).
For a state $q\in Q$, the first component $\pr_1(d(q))\in 2$ specifies the
finality of $q$, and the second component $\pr_2(d(q))\in Q^\Sigma$ 
sends input symbols $a\in \Sigma$ to successor states in $Q$ (the initial state $q_0 \in Q$ is not important here).

On the other hand, a non-deterministic automaton is simply a coalgebra for the
composed functor $H\Powf \colon \Set\to \Set$, i.e.~a map
\(
  c\colon Q\to 2\times (\Powf Q)^\Sigma
\).
The generalized determinization assumes that $T\colon \C\to \C$ is a monad (with unit
$\eta$ and multiplication $\mu$) and that
$H\colon \C\to\C$ is a functor that lifts to the Eilenberg\hyp{}Moore category
of $T$, i.e.~that we have a functor $G\colon \EM(T)\to \EM(T)$ with
$\phi\colon HU\xrightarrow{\cong} UG$ (\autoref{defLift}) -- for example
$HX = 2\times X^\Sigma$ and $TX=\Powf X$.
Then the generalized determinization turns an $HT$-coalgebra on $Q$ into a
$G$-coalgebra on $TQ$, using the adjunction $F\dashv U$ between $\C$ and
$\EM(T)$ (cf~\eqref{adjointEM}):
\[
  \begin{tikzcd}[arrow style=tikz,>={Straight Barb[scale=0.8]},]
  c\colon Q\to HTQ
  ~~\text{ in }\C\quad
  \arrow[mapsto,line width=1.4pt,draw=lipicsYellow,
  ]{r}{\textsf{Internalization}}
  &[2cm]
  ~~
  d\colon FQ\to GFQ
  ~~\text{ in }\EM(T)
  \end{tikzcd}
\]
In the instance $TX=\Powf X$ of the powerset construction, $\EM(T)$ is the category
of join-semilattices, and every non-deterministic automaton on $Q$ is turned
into a deterministic automaton on $\Powf Q$. Since we internalized the side-effect of non-deterministic branching in the states, the resulting state space $\Powf Q$ is the
configuration space of the non-deterministic automaton and the induced
transition structure $d$ preserves joins. Instantiating the generalized
determinization to supported sets and nominal sets yields a construction that internalizes the side-effect of rearranging atoms or registers.
Here, we stick to the equality symmetry $M:=\Perm(\A)$,
$TX:=\Monad[\Perm(\A)]X$, i.e.~we stick to \autoref{assCountable}.
\begin{construction} \label{nomGenDet}
  Fix $M:=\Perm(\A)$. For functors $H\colon \Supp{\A}\to\Supp{\A}$
  and $G\colon \Nom\to\Nom$ with $HU \cong UG$, we have the internalization process
  \[
    \begin{tikzcd}[arrow style=tikz,>={Straight Barb[scale=0.8]},]
      c\colon Q\to H(\Monad Q)
      ~\text{ in }\Supp{\A}~~
      \arrow[mapsto,line width=1.4pt,draw=lipicsYellow,
      ]{r}{{\text{\upshape \textsf{Internalization}}}}
      &[2cm]
      ~
      d\colon \Monad Q\to G(\Monad Q)
      ~\text{ in }\Nom.
    \end{tikzcd}
  \]
\end{construction}
\begin{proofappendix}[Details for]{nomGenDet}
  Apply the adjunction of the Eilenberg-Moore category
  \[
    \EM(\Monad(-))= \Nom
  \]
  to
  \[
    Q\xrightarrow{c} H(\Monad Q)
    \xrightarrow{\phi_Q} UG (\Monad Q).
  \]
  This yielding the desired equivariant map
  \[
    \Monad Q\xrightarrow{d} G (\Monad Q).
  \]
  If $Q$ is finitely presentable (cf.~\autoref{suppLFP}), then by
  \cite[Corollary~2.75]{adamek1994locally}, $\Monad Q$
  is finitely presentable in $\Nom$, i.e.~orbit-finite.
\end{proofappendix}
Here, the nominal set $\Monad Q$ is the configuration space (states + register
assignments) of the original automaton $(Q,c)$. The configuration $\id_A\cdot q$ in $\Monad Q$
behaves like $q\in Q$ in $c$, and the resulting transition structure $d$ is equivariant.
We can apply this construction to many different system types $H$ and $G$ that arise
from the following functors. Here, $\Powufs X$ (for a nominal set $X$) contains
those subsets $S\subseteq X$ for which $\bigcup\{\supp{}(x)\mid x\in S\}$ is
finite~\cite{skmw17}.
\begin{proposition}
  \label{liftClass}
  The functors $G$ on $\Nom$ that are the lifting of a functor $H$ on $\Supp{\A}$
  contain $\Powf$, $\Powufs$, $\Abs$, and all constant functors and are closed under
  all (possibly infinite) products and coproducts.
\end{proposition}
\begin{proofappendix}{liftClass}
  \begin{definition}
    The \emph{uniformly supported powerset functor} $\Powufs\colon \Nom\to\Nom$ is given 
    by
    \[
      \Powufs (X) =\set{
        S\subseteq X
        \mid
        \bigcup_{x\in S}\supp{X}(x)\text{ is finite}
      }
    \]
    On equivariant maps $f$, $\Powufs f$ performs the direct image.
  \end{definition}
  For the verification of \autoref{liftClass}:

  \begin{itemize}
  \item Finite powerset $\Powf\colon \Nom\to\Nom$ has precisely the same
    definition in $\Nom$ and $\Supp{\A}$. In fact, they both lift from $\Powf$
    on $\Set$.
  \item The uniformly supported powerset functor on $\Nom$ is a lifting of the
    obvious functor $\Powufs'$on $\Supp{\A}$:
    \begin{align*}
    \Powufs'&\colon \Supp{\A}\to \Supp{\A}
              \\
      \Powufs'&(X) = \set{
                S\subseteq X
                \mid
                \bigcup_{x\in S}\s{X}(x)\text{ is finite}
                }
      \\
      &\s{\Powufs'(X)}(S) = \bigcup_{x\in S}\s{X}(x)
    \end{align*}
    Since $\s{UX}(x) = \supp{X}(x)$ for all nominal sets $X$, we have that
    \[
      \Powufs' UX = U\Powufs X
    \]
  \item Every constant functor $C\colon \Nom\to \Nom$, $CX=N$ is the lifting of the 
    constant functor $C'\colon \Supp{\A}\to\Supp{\A}$, $C'X= UN$.
  \item If the family $G_i\colon \Nom\to\Nom$, $i\in I$ are liftings of the
    functors $H_i\colon \Supp{\A}\to\Supp{\A}$, then
    \begin{itemize}
    \item $\prod_{i\in I} G_i$ is the lifting of $\prod_{i\in I} H_i$, because
      the right-adjoint $U\colon \Nom\to\Supp{\A}$ preserves products:
      \[
        U\prod_{i\in I} G_i
        \cong \prod_{i\in I} UG_i
        \cong \prod_{i\in I} H_iU
      \]
    \item Coproducts are given by disjoint union, both in $\Nom$ and in $\Supp{\A}$. Hence,
      $U\colon \Nom\to\Supp{\A}$ preserves coproducts and so
      $\coprod_{i\in I} G_i$ is the lifting of $\coprod_{i\in I} H_i$:
      \[
        U\coprod_{i\in I} G_i
        \cong \coprod_{i\in I} UG_i
        \cong \coprod_{i\in I} H_iU
        \tag*{\qedhere}
      \]
    \end{itemize}
  \end{itemize}
\end{proofappendix}

\begin{example}
  All $\Nom$-functors arising from binding
  signatures~\cite{nomcoalgdata,FiorePlotkinTuri99} are such liftings. In particular,
  \(
    \Lambda X = \A + \Abs X + X\times X
  \)
  (\autoref{exLambdaFunctor})
  is the lifting of
  \(
    H X = \A + \B X + X\times X
  \).
  The coalgebras of $\Lambda$ are possibly infinite $\lambda$-trees modulo
  $\alpha$-equivalence~\cite{nomcoalgdata}. Such a $\lambda$-tree can then be represented by a
  supported set $X$ and a supported map \[
    f\colon X\to \A ~+~ \B (\Monad[\Perm(\A)] X) ~+~ (\Monad[\Perm(\A)] X)\times (\Monad[\Perm(\A)] X)
  \]
\end{example}
\begin{example}
  Nominal automata with name binding considered by
  Schröder\ \etal\ \cite{skmw17} are coalgebras for the $\Nom$-functor
  \(
    KX = 2\times \Powufs(\Abs X+ \A\times X)
  \)
  which is a lifting of $HX = 2\times \Powufs(\B X+ \A\times X)$ on supported
  sets. Thus, these nominal automata can be represented by finite coalgebras in $\Supp{\A}$.
\end{example}

Interpreting $\A$ as the names of registers, register automata in the style of
Cassel~\etal~\cite{CasselHJS16}
 straightforwardly adapt to $HT$-coalgebras
in supported sets:
\begin{example}\label{exRegAut}
  Let $\G$ be a nominal set of \emph{guards},
  where we understand $g\in \G$ as a predicate involving register names $\supp{G}(g)\in \Powf(\A)$,
  e.g.~$\mathsf{iszero}(a)$, $\mathsf{divides}(a, b)$, $\mathsf{plus}(a, b, c)$.
  It is important that the atoms $a\in \A$ are the register names, and not the
  data itself. So e.g.~$\mathsf{iszero}(a_3)$ is satisfied if the data value stored
  in register $a_3$ is zero -- this is the \emph{symbolic} semantics of
  register automata~\cite{VM20Ictac,VaandragerM22}.
  Using the forgetful $U\colon \Nom\to \Supp{\A}$, we can also use $\G$ as a supported set where
  the support of $g\in \G$ is the set of register names $a\in \A$ appearing in $g$.
  Now, a register automaton is a $HT$-coalgebra in $\Supp{A}$
  \[
    c\colon Q\rightarrow 2\times \B \Powf(U\G\times \Monad[\Perm(\A)]Q)
    \qquad\text{for}~~
    HX= 2\times \B \Powf(\G\times X)
    ~~\text{and}~~
    TX=\Monad[\Perm(\A)] X.
  \]
  As before, the first component of $c(q)$ defines the finality of a state $q\in
  Q$. While in state $q$, the registers $\s{}(q)\subseteq \A$ are filled with
  data and the second component of $c(q)$ binds the next input data symbol into a
  fresh register ($\B$) and then provides a finite number of transitions
  of the form $q\xrightarrow{g,\pi} q'$.
  In such a transition, the guard $g\in \G$, specifies when this transition can
  be taken, depending on the register contents of $q$ and the freshly read input
  symbol. When a transition is taken, the registers are rearranged via the
  permutation $\pi$ before state $q'$ is entered.
  The map $\pi\colon \s{}(q')\monoto \A$ specifies for each register $r\in \s{}(q')$ 
  of the target set, where the data symbol is drawn from. The map $c$ being a supported
  map ensures that whenever we transfer to state $q'$, then all registers
  $\s{}(q')$ can be filled with data, coming for the support of the previous
  state $q$ or from the \textqt{input register} in $\B$ (a concrete register
  automaton is discussed in \autoref{exExRA}). The internalization process
  turns $c$ into an equivariant map $d\colon \bar Q\to 2\times \Abs
  \Powf(\G\times \bar Q)$, which is the nominal automaton of configurations
  $\bar Q = \Monad[\Perm(\A)] Q$.

  The construction also nicely interacts with \emph{initial} states. An initial
  state of a register automaton is simply a supported map $i\colon 1\to Q$, that is, an
  element of $Q$ with empty support. Applying the functor $\Monad[\Perm(\A)]$
  to $i$ yields an equivariant map, $\Monad[\Perm(\A)]i\colon
  \Monad[\Perm(\A)]1\to \bar Q$. Since the only element of $1=\{0\}$ has empty support, we have
  $\Monad[\Perm(\A)]1 \cong 1$, so $\Monad[\Perm(\A)]i$ is equivalent to an
  equivariant map $i'\colon 1\to \bar Q$.
\end{example}
\begin{proofappendix}[Details for]{exRegAut}
  In \autoref{exRegAut}, we simplified the presentation a lot, especially the
  definition of guards; the full definitions are as follows.

  A \emph{(relational) signature} is a set $\R$ with an arity map $\ar\colon \R\to \N$;
  The nominal set of \emph{guards} $\G$ over this relational signature $\R$ is
  \[
    \G = \big(2\times \coprod_{r\in \R} \A^{\ar(r)}\big)^*
  \]
  Intuitively, the words $(-)^*$ represent conjunction, $2$ encodes a
  possible negation, the coproduct over $\R$ is the set of relational terms over $\A$.
  Hence, a guard $g\in \G$ is a finite conjunction of
  possibly negated terms of the form $r(a_1,\ldots,a_n)$, $n=\ar(r)$,
  $a_1,\ldots,a_n\in \A$. Hence, the support of a guard $g\in \G$ is the set of
  register names $a\in \A$ that appear in it. Then, a register
  automaton~(in the style of \cite{CasselHJS16}) for $\R$ is a tuple $(Q,q_0,f,\Gamma)$ where
  \begin{enumerate}
  \item $Q$ is a finite supported set of \emph{locations}
  \item $q_0\colon 1\to Q$ is a supported map, the \emph{initial location},
  \item $f\colon Q\to 2$ is a supported map, indicating \emph{finality},
  \item $\Gamma\colon Q\to \B \Powf(\G\times \Monad[\Perm(\A)] Q)$ is a
    supported map, the \emph{transitions}.
  \end{enumerate}
  Here, the properties of supported sets and maps encode precisely the coherence
  conditions of register automata:
  \begin{enumerate}
  \item Every location has access to finitely many registers.
  \item The initial location has all registers uninitialized.
  \item There is no side condition on $f$ since 2 is a set.
  \item Every transition $q\xrightarrow{g,\pi} q'$ is in such a way that:
    \begin{itemize}
    \item the guard $g$ may only use registers from $\s{}(q)$ and the
      input data value.
    \item $\pi$ tells for each register of $q'$, $\pi$ its value (from the
      previous register contents or the input data value).
    \end{itemize}
    If the register contents and the input value satisfy the guard $g$, then $q$
    makes a transition to location $q'$ with the registers rearranged
    with respect to the injective map
    \[
      \pi\colon \s{Q}(q')\monoto \{\rho(0)\} \cup \{ \myold(a)\mid a\in \s{Q}(q)\}
    \]
    where $\rho(0)$ is the input value and $\myold(\rho(k)) = \rho(k+1)$ refers
    to the old registers contents, i.e.~$\myold(\rho(k))$
    has the role that $\rho(k)$ had above the binder $\B$.
  \end{enumerate}
  
  In one line, we have an $HT$-coalgebra on $Q$ for
  \[
    HX= 2\times \B \Powf(\G\times X)
    \quad\text{and}\quad
    TX=\Monad[\Perm(\A)] X.
  \]
  Since $\G$ is a nominal set, we can consider it as a constant functor
  $C_\G\colon \Nom\to \Nom$, $C_\G(X) = \G$. This functor is the
  lifting of the constant $C_{U\G}\colon \Supp{\A}\to\Supp{\A}$, $C_{U\G}(X) =
  U\G$ because $U\circ C_{\G} = C_{U\G}\circ U$.
  By \autoref{liftClass}, $H$ lifts to a functor
  \[
    G\colon \Nom\to\Nom
    \qquad
    G(X) = 2\times \Abs\Powf(\G\times X)
  \]
  and
  so \autoref{nomGenDet} transforms a register automaton into a nominal set
  $\bar Q$
  and an equivariant map
  \[
    d\colon
    \bar Q
    \longrightarrow
    2\times \Abs \Powf(\G\times \bar Q)
  \]
  that is, a nominal automaton.
\end{proofappendix}
\begin{example}\label{exExRA}%
  \begin{figure}%
    \hspace*{1cm}
    \newcommand{\mathsplitnode}[3]{%
  \begin{scope}[on background layer]
    \draw[draw=lipicsYellow,line width=1pt] (#1.west) -- (#1.east);
  \end{scope}
  \begin{scope}[every node/.append style={
    inner sep=2pt,
  }]
    \node[anchor=south] at (#1.center) {\ensuremath{#2}};
    \node[anchor=north] at (#1.center) {\ensuremath{\set{#3}}};
  \end{scope}
}%
\begin{tikzpicture}[
    state/.style={
      draw=lipicsYellow,
      line width=1pt,
      shape=circle,
      align=center,
      inner sep=0pt,
      minimum size=12mm,
      double,
      double distance=1pt,
    },
    transition/.style={
      -{Straight Barb[scale=0.8]},
      draw=lipicsGray,
      line width=1pt,
      shorten >= 2pt,
      shorten <= 2pt,
      every node/.append style={
        align=center,
        font=\footnotesize,
      },
    },
    x=3cm,
  ]
  \node[state] (q0) at (0,0) {}; \mathsplitnode{q0}{q_0}{};
  \node[state] (q1) at (1,0) {}; \mathsplitnode{q1}{q_1}{\ell};
  \node[state] (q2) at (2,0) {}; \mathsplitnode{q2}{q_2}{\hspace*{-1pt}\ell,s\hspace*{-1pt}};
  \path[transition] ([xshift=-5mm]q0.west) -- (q0);
  \path[transition,bend left=20] (q0) edge
    node[above] {$\top$}
    node[below] {$\ell \mapsto p$}
    (q1);
  \path[transition,bend left=20] (q1) edge
    node[above] {$\top$}
    node[below] {$\begin{array}{r@{\,}l}
      s &\mapsto \old(\ell) \\
      \ell &\mapsto p
      \end{array}$}
    (q2);
  \path[transition] (q2) edge[overlay,loop,out=-30,in=30,looseness=6]
    node[anchor=west] {\(\plus(s,\ell,p)\)\\[1mm]
      $\begin{array}{r@{\,}l}
      s &\mapsto \old(\ell) \\
      \ell &\mapsto p
      \end{array}$
    }
  (q2);
\end{tikzpicture}%
    \caption{Example (symbolic) register automaton}
    \label{figExRA}
  \end{figure}%
  An example register automaton is visualized in \autoref{figExRA}: the upper
  half of the nodes provide the state names $Q=\set{q_0,q_1,q_2}$, the lower
  half specifies their support $\s{Q}(q_0) = \emptyset$, $\s{Q}(q_1) =
  \set{\ell}$, $\s{Q}(q_2) = \set{s}$, where $\ell,s\in \A, \ell\neq s$ are
  arbitrary (standing for $\ell$ast and $s$econd last). The initial state is $q_0$, i.e.~the supported map $i\colon 1\to
  Q$ is $i(0) = q_0$, and all states are final.
  We mimic existing register automata notation~\cite{CasselHJS16,VM20Ictac}
  by defining $p:= \rho(0)\in \A$ (note that we do \emph{not} require $\ell,s$ to
  be distinct from $p$). Also, we use $\old\colon \A\to\A$ defined by
  $\old(\rho(k)) = \rho(k+1)$, which satisfies $a\in \s{\B X}(\bind x)$ iff
  $\old(a) \in \s{X}(x)$ for all supported sets $X$ and $x\in X$.
  The nominal set of guards is defined by
  \[
    \G := \set{\top} + \set{\plus}\times \A^3,
  \]
  so the constant $\top$ represents \textqt{true} and the ternary relation
  symbol $\plus(a,b,c)$ represents that the sum of the register contents of $a$
  and $b$ equals the register content of $c$. Since we use symbolic semantics,
  the concrete data domain does not need to be specified here; but of course
  one can interpret $\plus$ over rational numbers for example.
  The supported map $c\colon Q\to 2\times \B\Powf(U\G\times \Perm[\A] Q)$ for
  transitions in the automaton is sincerely visualized in \autoref{figExRA}:
  \begin{itemize}
  \item The transition $q_0\xrightarrow{g,\pi}q_1$ has the guard $g=\top$ and
  the register reassignment $\pi\colon \s{}(q_1)\monoto \A$ is defined by
  $\pi(\ell) = p$, meaning that when entering state $q_1$, the register $\ell$
  will be filled with what was in the input $p$ before:
  $c(q_0) = (1, \bind\set{(\top, (\pi, q_1))})$.
  This satisfies support preservation because $\s{}(\top, (\pi, q_1)) = \pi[\s{}(q_1)] = \set{p}$
  and so $\s{}(\bind(\top, (\pi, q_1))) = \emptyset\subseteq \s{}(q_0)$.

  \item The transition $q_1\xrightarrow{g,\sigma}q_2$ has the same guard
  $g=\top$ and $\sigma\colon \s{}(q_2)\monoto \A$ is defined by $\sigma(s) =
  \old(\ell)$ and $\sigma(\ell) = p$.
  Again, $c(q_1) = (1, \bind\set{(\top, (\sigma, q_2))})$ preserves support because
  \[
    \s{}(\top, (\sigma, q_2)) = \sigma[\s{}(q_2)] = \sigma[\set{\ell,s}]
    = \set{p,\old(\ell)}
    \quad
    \text{and}
    \quad
    \s{}(\bind\set{(\top, (\sigma, q_2))}) = \set{\ell}
    \subseteq \s{}(q_1).
  \]
  \item The loop $q_2\xrightarrow{g',\sigma}q_2$ has the guard
  $g'=\plus(s,\ell,p)$ and literally the same $\sigma\colon \s{}(q_2)\monoto \A$ as
  in the previous transition. Support preservation holds for the mapping $c(q_1) = (1, \bind\set{(\top, (\sigma,
  q_2))})$, because
  \[
     \s{}(\plus(s,\ell,p), (\sigma, q_2)) = \set{\old(s),\old(\ell),p}
     ~\text{and}~
    \s{}(\bind\set{(\plus(s,\ell,p), (\sigma, q_2))}) = \set{s,\ell}
    \subseteq \s{}(q_2).
  \]
  \end{itemize}
  The coherence axioms of register automata naturally translate into $c$ being a supported map.
  \autoref{nomGenDet} transforms this finite coalgebra in $\Supp{\A}$ into a nominal
  automaton, in the style of symbolic semantics~\cite{VM20Ictac} of register automata.
\end{example}

\section{Conclusions and Future Work}
We have seen that by going from the base category of sets to supported sets,
nominal sets for various symmetries surprisingly turn out to be monadic.
Supported sets have a functor for name binding, which
even lifts to the abstraction functor in nominal sets. It remains for future
work whether a similar name binding functor can be found for other data
alphabets, most notably for the total order symmetry on $\Q$, and whether
multiple atoms can be bound simultaneously, as it is possible in nominal
sets~\cite{Clouston13}.
It can be conjectured that such generalizations are not possible on the level
of supported sets.

On the positive side, due to the little structure of supported sets, it provides
a common foundation for the nominal sets for different symmetries, in the sense
of being described by monads on supported sets. The monadicity can be used to relate nominal automata with
register automata, which have a natural definition in supported sets. It remains
for future investigation how the \emph{data semantics} of register automata can be
phrased in supported sets. We are optimistic that it helps to develop a categorical
semantics for register automata for data alphabets and signatures beyond symmetries (e.g.~those
for priority queues \cite{CasselHJS16}). When developing algorithms, in particular
learning algorithms and minimization algorithms, for register or nominal automata~\cite{CasselHJS16,DBLP:conf/lics/UrbatS20,DBLP:conf/popl/MoermanS0KS17}, supported sets directly yield a finite
representation that can help in the implementation and complexity analysis.

\label{maintextend} %
\bibliographystyle{plainurl}
\bibliography{refs}

\vfill
\pagebreak
\appendix

\ifthenelse{\boolean{proofsinappendix}}{%
\section{Omitted Proofs and Further Details}
\closeoutputstream{proofstream}
\renewcommand{\labelmarginpar}[1]{}
\input{\jobname-proofs.out}
}{%
}

\end{document}
